\documentclass[twocolumn,showpacs,superscriptaddress,aps,prc,10pt,nofootinbib]{revtex4-1}
\usepackage[pdftex]{graphicx}
\usepackage[usenames,svgnames,table]{xcolor}
\usepackage[english]{babel}
\usepackage{ucs}
\usepackage{tabularx} % in the preamble
\usepackage[utf8x]{inputenc}
\usepackage{siunitx}
\usepackage[version=3]{mhchem}
\usepackage{multirow}
\usepackage{soul}
\newcommand{\faz}{\textsc{FAZIA}}

\newcommand{\cb}{$^{48}$Ca}

\begin{document}
\sisetup{exponent-product = \cdot,per-mode = symbol,range-phrase = --,list-units = single,range-units = single,separate-uncertainty = true,table-number-alignment = center}

\preprint{}

\title{Characterization of the breakup channel in the asymmetric systems $^{40,48}$Ca+$^{12}$C at 25 and 40 MeV/nucleon}
% Force line breaks with \\

 \author{S.~Piantelli}
 \email{Corresponding author. e-mail: silvia.piantelli@fi.infn.it}
 \affiliation{INFN Sezione di Firenze, I-50019 Sesto Fiorentino, Italy}
 \author{G.~Casini}
 \affiliation{INFN Sezione di Firenze, I-50019 Sesto Fiorentino, Italy} 
\author{P.~Ottanelli}
 \affiliation{INFN Sezione di Firenze, I-50019 Sesto Fiorentino, Italy} 
\author{L.~Baldesi}
 \affiliation{Dipartimento di Fisica, Universit\`a di Firenze, I-50019 Sesto Fiorentino, Italy} 
 \affiliation{INFN Sezione di Firenze, I-50019 Sesto Fiorentino, Italy} 
\author{S.~Barlini}
\affiliation{Dipartimento di Fisica, Universit\`a di Firenze, I-50019 Sesto Fiorentino, Italy} 
 \affiliation{INFN Sezione di Firenze, I-50019 Sesto Fiorentino, Italy} 
\author{B.~Borderie}
 \affiliation{Universit\'e Paris-Saclay, CNRS/IN2P3, IJCLab, 91405 Orsay, France}
\author{R.~Bougault}
 \affiliation{Normandie Universit\'e, ENSICAEN, UNICAEN, CNRS/IN2P3, LPC Caen, 14000 Caen, France} 
\author{A.~Camaiani}
\email{present address: Dipartimento di Fisica, Universit\`a di Firenze, I-50019 Sesto Fiorentino, Italy} 
\affiliation{Instituut voor Kern-en Stralingsfysica, KU Leuven, B-3001 Heverlee, Belgium} 
\author{A.~Chbihi}
\affiliation{GANIL, CEA/DRF-CNRS/IN2P3, 14076 Caen, France} 
\author{C.~Ciampi}
\email{present address: GANIL, CEA/DRF-CNRS/IN2P3, 14076 Caen, France} 
 \affiliation{Dipartimento di Fisica, Universit\`a di Firenze, I-50019 Sesto Fiorentino, Italy} 
 \affiliation{INFN Sezione di Firenze, I-50019 Sesto Fiorentino, Italy} 
\author{J.~A.~Due$\tilde{\mathrm{n}}$as}
 \affiliation{Departamento de Ingenier\'a El\'ectrica y Centro de Estudios Avanzados en F\'isica, Matem\'aticas y Computaci\'on,
Universidad de Huelva, 21007 Huelva, Spain} 
\author{D.~Fabris}
 \affiliation{INFN Sezione di Padova, 35131 Padova, Italy} 
\author{Q.~Fable}
\affiliation{Laboratoire des 2 Infinis - Toulouse (L2IT-IN2P3), Universit\'e de Toulouse, CNRS, UPS, F-31062 Toulouse Cedex 9, France}
\author{J.~D.~Frankland}
\affiliation{GANIL, CEA/DRF-CNRS/IN2P3, 14076 Caen, France}
\author{C.~Frosin}
 \affiliation{Dipartimento di Fisica, Universit\`a di Firenze, I-50019 Sesto Fiorentino, Italy} 
 \affiliation{INFN Sezione di Firenze, I-50019 Sesto Fiorentino, Italy} 
\author{F.~Gramegna}
\affiliation{INFN Laboratori Nazionali di Legnaro, 35020 Legnaro, Italy}
\author{D.~Gruyer}
\affiliation{Normandie Universit\'e, ENSICAEN, UNICAEN, CNRS/IN2P3, LPC Caen, 14000 Caen, France} 
\author{B.~Hong}
 \affiliation{ Center for Extreme Nuclear Matters (CENuM), Korea University, Seoul 02841, Republic of Korea}
 \affiliation{ Department of Physics, Korea University, Seoul 02841, Republic of Korea}
\author{A.~Kordyasz}
 \affiliation{Heavy Ion Laboratory, University of Warsaw, 02-093 Warszawa, Poland}
\author{T.~Kozik}
 \affiliation{Faculty of Physics, Astronomy and Applied Computer Science, Jagiellonian University, 30-348 Krak\'ow, Poland}
\author{M.~J.~Kweon}
\affiliation{Department of Physics, Inha University, Incheon 22212, Republic of Korea}
\author{J.~Lemari\'e}
\affiliation{GANIL, CEA/DRF-CNRS/IN2P3, 14076 Caen, France}
\author{N.~LeNeindre}
 \affiliation{Normandie Universit\'e, ENSICAEN, UNICAEN, CNRS/IN2P3, LPC Caen, 14000 Caen, France} 
\author{I.~Lombardo}
 \affiliation{INFN Sezione di Catania, 95123 Catania, Italy} 
 \affiliation{Dipartimento di Fisica e Astronomia, Universit\`a di Catania, 95123 Catania, Italy} 
\author{O.~Lopez}
\affiliation{Normandie Universit\'e, ENSICAEN, UNICAEN, CNRS/IN2P3, LPC Caen, 14000 Caen, France}
\author{T.~Marchi}
\affiliation{INFN Laboratori Nazionali di Legnaro, 35020 Legnaro, Italy}
\author{K.~Mazurek}
\affiliation{Institute of Nuclear Physics Polish Academy of Sciences, PL-31342 Krak\'ow, Poland}
\author{S.~H.~Nam}
\affiliation{ Center for Extreme Nuclear Matters (CENuM), Korea University, Seoul 02841, Republic of Korea}
\affiliation{ Department of Physics, Korea University, Seoul 02841, Republic of Korea}
\author{M.~P$\hat{\mathrm{a}}$rlog}
 \affiliation{Normandie Universit\'e, ENSICAEN, UNICAEN, CNRS/IN2P3, LPC Caen, 14000 Caen, France}
 \affiliation{"Horia Hulubei" National Institute of Physics and Nuclear Engineering (IFIN-HH), RO-077125 Bucharest Magurele, Romania}
\author{J.~Park}
\affiliation{ Center for Extreme Nuclear Matters (CENuM), Korea University, Seoul 02841, Republic of Korea}
\affiliation{ Department of Physics, Korea University, Seoul 02841, Republic of Korea}
\author{G.~Pasquali}
 \affiliation{Dipartimento di Fisica, Universit\`a di Firenze, I-50019 Sesto Fiorentino, Italy} 
 \affiliation{INFN Sezione di Firenze, I-50019 Sesto Fiorentino, Italy} 
\author{G.~Poggi}
 \affiliation{Dipartimento di Fisica, Universit\`a di Firenze, I-50019 Sesto Fiorentino, Italy} 
 \affiliation{INFN Sezione di Firenze, I-50019 Sesto Fiorentino, Italy} 
\author{A.~Rebillard-Souli\'e}
 \affiliation{Normandie Universit\'e, ENSICAEN, UNICAEN, CNRS/IN2P3, LPC Caen, 14000 Caen, France} 
\author{A.~A.~Stefanini}
 \affiliation{Dipartimento di Fisica, Universit\`a di Firenze, I-50019 Sesto Fiorentino, Italy} 
 \affiliation{INFN Sezione di Firenze, I-50019 Sesto Fiorentino, Italy} 
\author{S.~Upadhyaya}
 \affiliation{Faculty of Physics, Astronomy and Applied Computer Science, Jagiellonian University, 30-348  Krak\'ow, Poland}
\author{S.~Valdr\'e}
 \affiliation{INFN Sezione di Firenze, I-50019 Sesto Fiorentino, Italy}
\author{G.~Verde}
 \affiliation{INFN Sezione di Catania, 95123 Catania, Italy}
 \affiliation{Laboratoire des 2 Infinis - Toulouse (L2IT-IN2P3), Universit\'e de Toulouse, CNRS, UPS, F-31062 Toulouse Cedex 9, France}
\author{E.~Vient}
\affiliation{Normandie Universit\'e, ENSICAEN, UNICAEN, CNRS/IN2P3, LPC Caen, 14000 Caen, France}
\author{M.~Vigilante}
\affiliation{Dipartimento di Fisica, Universit\`a di Napoli, 80126 Napoli, Italy} 
\affiliation{INFN Sezione di Napoli, 80126 Napoli, Italy}
\collaboration{FAZIA Collaboration}
\noaffiliation

\date{\today}% It is always \today, today,

\begin{abstract}
An analysis of the asymmetric reactions $^{40,48}$Ca+$^{12}$C at 25
and 40 MeV/nucleon is presented. Data have been collected with six modules
of the FAZIA array. The analysis is focused on the breakup
channel of sources produced in  dissipative collisions, partially corresponding to
incomplete fusion processes.
The study has been performed both on detected fragments
and on some resonances reconstructed by means of particle-fragment
correlations, with a focus on the evolution of the breakup channel with the beam energy and
the neutron content of the system, looking in particular at the relative velocity between the breakup fragments. Results show that also
 Carbon fragments reconstructed by means 
of particle  correlations can be in large part interpreted as the light partner of a scission. 

\end{abstract}

%Use showkeys class option if keyword
\maketitle

\section{Introduction}
\label{introd}
The investigation of the breakup channel in heavy ion collisions at
Fermi energies
is a topic widely represented in the literature
since many years. In particular, efforts have been done to  study the
emission pattern of the fragments coming from the QuasiProjectile (QP)
breakup
\cite{Casini93,gingras_prc65_2002,DeFilippo05,DeFilippo2012,defilpaga2014,Jedele2017,Manso2017,hannaman_prc2020,Camaiani2021,Piantelli2020}. For example, one goal is the investigation of the differences between this process and a pure low-energy fission modelized
in terms of the nuclear structure and statistical arguments. In
particular, a correlation between  
 the scission timescales, the 
breakup configuration and the split asymmetry has been found: for
large mass asymmetry, the lighter partner is preferentially emitted in
the backward direction and on fast times, therefore this process was
called fast oriented fission \cite{Casini93,Bocage2000,DeFilippo05,McIntosh2010}. More recently, very refined
investigations have been carried out~\cite{Jedele2017,Manso2017} also by our collaboration~\cite{Piantelli2020,Camaiani2021}  concerning  the isotopic composition
of both QP breakup fragments in
coincidence.
In particular, in
\cite{Jedele2017,Manso2017} the 
isotopic composition of the breakup partners (with more sensitivity
for the lighter one)  has been found to be correlated with 
the breakup configuration: when the light fragment is backward emitted
towards the center of mass (c.m.) of the system, i.e. in  configurations typical of
the fast oriented fission, it is more neutron
rich than when emitted in the opposite
direction. According to the general consensus on this
subject~\cite{defilpaga2014,DeFilippo05}, also recently reinforced by
theoretical analyses~\cite{harvey_prc102_2020}, 
the authors of the papers~\cite{Jedele2017,Manso2017} interpreted the
angle  between the QP breakup axis and the QP-QT (QuasiTarget) separation axis 
(often labeled as $\alpha$ or $\theta_{PROX}$)
as a clock of the fission process and they extracted a timescale of
about 100~fm/c for the
equilibration of the isospin degree of freedom within the QP. 
In \cite{Piantelli2020}, 
in the limit of the available statistics, a similar analysis was performed,
finding a correlation  of the
isospin versus angle $\alpha$ in
substantial agreement with the results of 
\cite{Jedele2017,Manso2017}. The experimental data were also found
quite consistent with the predictions
of the antisymmetrized molecular dynamics model
(AMD)~\cite{Ono92,Ono99,Ono02,OnoNN2012,OnoJPC2013}.
However, the
interpretation of the $\alpha$ angle as a clock of the fission
process (and of the isospin equilibration)
was not obvious within the framework of AMD.

The above discussion on the breakup pertains to rather symmetric
collisions between relatively heavy nuclei that are dominated 
by a large
yield of almost binary events with   two main 
QP and  
QT  ejectiles; as discussed above, these deformed and excited
nuclei can then undergo 
a  breakup process with different mass asymmetries possibly
associated with various timescales.
Obviously, at Fermi beam energies (i.e. in the range 20-50
MeV/nucleon) a  fission-like process can occur not
  only for QP and/or QT 
in semiperipheral
collisions but also for composite
sources produced in incomplete fusion events.
Fusion-like events are a minority in case of heavy systems but they can 
generally be favoured moving to lighter and more
asymmetric reactions \cite{Morgenstern}. 
For mass asymmetric systems, the contribution of binary dissipative collisions 
is confined at rather peripheral reactions. At energies just above the
  Coulomb barrier,
reactions of the 
fusion type dominate at almost all the impact parameters, with the
formation of a unique 
major excited source. With increasing  beam energy, fast emissions of
nucleons or clusters  during the interaction (before the
formation of a ``true'' compound nucleus)  become more important  thus
favouring the incomplete fusion channel. The rest of
the reaction cross section for asymmetric medium-light systems
and at Fermi energies, corresponds to a broad distribution of more or
less damped collisions  with
the formation of  excited sources  which further break up and decay.
Of course, the distinction between very dissipative binary reactions
and incomplete fusion is faint; in particular for asymmetric systems,
also taking into account the broadening due to fast emissions, it is
difficult to 
disentangle the sources produced through
incomplete fusions or through damped (binary) reactions by using the measured products.
Indeed, from the experimental point of view, the overlap of concurrent
processes  
generate a broad spectrum of primary sources which then decay towards
stable states; in reverse kinematics
measurements, as often done, this corresponds to the observation at
forward angles of a wide range of fragments with sizes not too far
from the projectile and velocities extending from those of the beam to
that of the c.m..
The experimental
data discussed in this paper refer to reactions of this  kind.

To our knowledge, medium-light asymmetric systems have not been widely studied,
in recent years, with reference to the characterization of the breakup
channel. Indeed, the reactions discussed  
in this paper, using Ca-beams of
different isotopes, provide also information about the
role of the neutron richness of the system on the 
source breakup. 
In
fact, this work deals with the analysis of experimental data collected
with six  \faz ~blocks  for the systems $^{48,40}$Ca+$^{12}$C
at 25 MeV/nucleon ~and $^{48}$Ca+$^{12}$C at 40 MeV/nucleon ~(referred to as 
FAZIAPRE experiment). The main focus of the paper is on the breakup channel 
for which we were able to attempt detailed analysis also benefiting of the
isotopic identification of both breakup partners. In particular we
investigated their relative velocity as a function of the charge
asymmetry of the splitting and of the reconstructed source size.  
Moreover, the particle correlation technique has been applied to
reconstruct some resonances of Carbon isotopes,
finding that also these excited fragments can be interpreted as  
light ejectiles of a breakup process.

\section{The experiment}

\begin{table*}[htbp]
\begin{tabular}{|c|c|c|c|c|c|c|c|}
\hline
Beam&(N/Z)$_{proj}$&(N/Z)$_{sys}$&$\vartheta_{graz}$&b$_{graz}
$&$v_{c.m.}$&$v_{beam}$&$v_{beam}^{c.m.}$\\
MeV/nucleon&&&deg&fm&mm/ns&mm/ns&mm/ns\\
\hline\hline
$^{40}$Ca@25&1.0&1.0&1.0&8.6&53.5&69.5&16.0\\
\hline
$^{48}$Ca@25&1.4&1.31&0.9&8.8&55.6&69.5&13.9\\
\hline
$^{48}$Ca@40&1.4&1.31&0.5&8.9&70.3&87.9&17.6\\
\hline
\end{tabular}
\caption{Some characteristics of the investigated reactions. (N/Z)$_{proj}$ and (N/Z)$_{sys}$ are the isotopic composition of the projectile and of the whole system, respectively. $\vartheta_{graz}$ and b$_{graz}$ are the grazing angle in lab and the grazing impact parameter, respectively. $v_{c.m.}$, $v_{beam}$ and $v^{c.m.}_{beam}$ are the c.m. velocity, the beam velocity in lab and in c.m., respectively.}
\label{tab1}
\end{table*} 

\begin{figure}[htbp]
\includegraphics[width=0.5\textwidth]{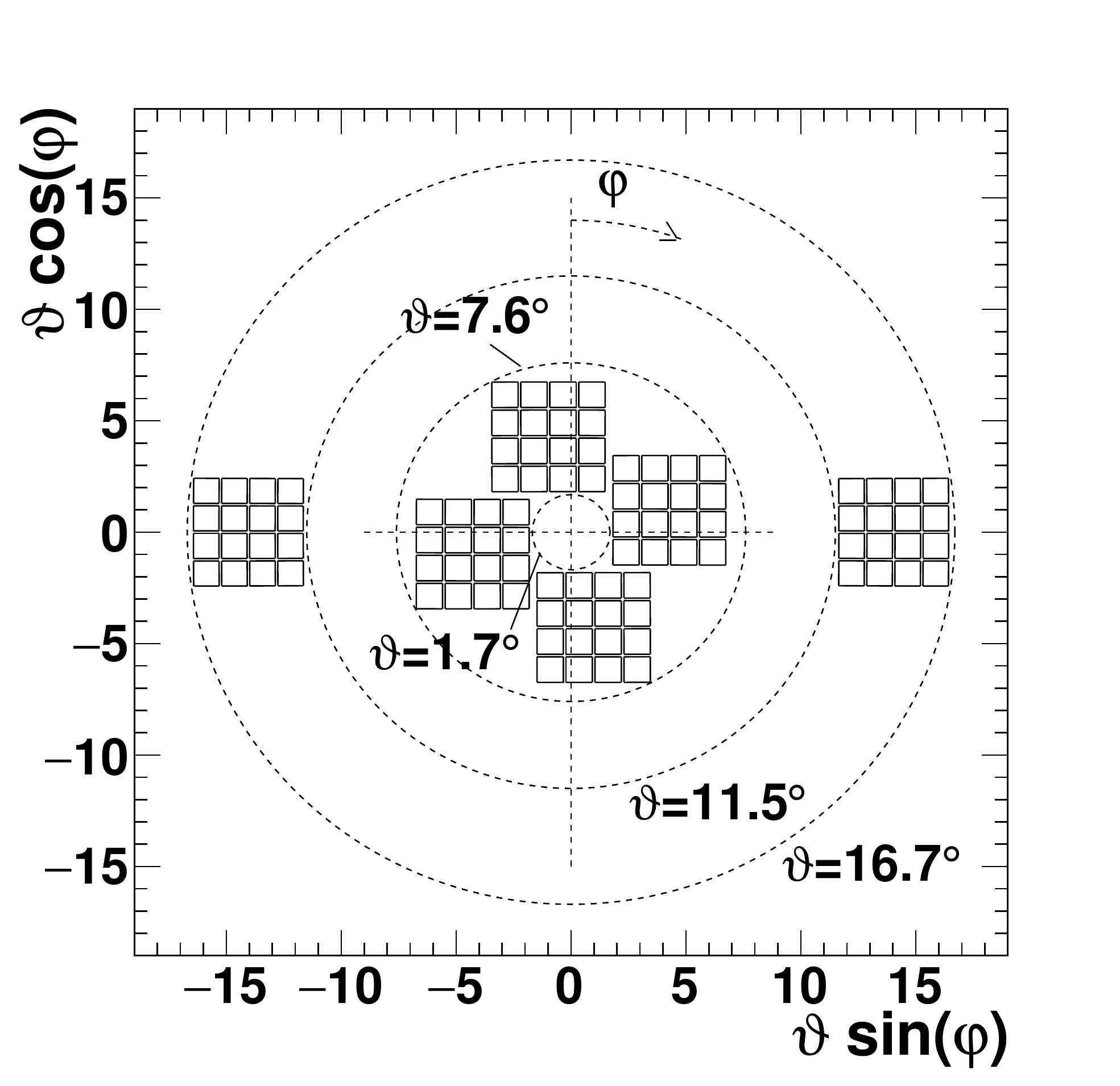}\\
\caption{Layout of the experimental setup in polar representation.}
\label{fig0}
\end{figure}

Pulsed beams of $^{48,40}$Ca at 25 and 40 MeV/nucleon, with average
current of 0.1 pnA, were delivered by the CS cyclotron of
INFN-LNS in Catania and impinged on a $^{12}$C target with a thickness of
about 300 $\mu$g cm$^{-2}$. The system $^{40}$Ca at 40
MeV/nucleon presents some calibration issues and it is not included
in the following analysis.    
Data were collected with six FAZIA blocks, four of them placed in a
wall configuration (1 m far from the target) and covering
the polar angles between 1.7$^{\circ}$ and 7.6$^{\circ}$ in the lab
frame, and two of them located on the equatorial plane on opposite
sides (polar angle range: 
11.5$^{\circ}$-16.7$^{\circ}$). 
The layout of the experimental setup in polar representation is shown
in Fig.~\ref{fig0} while
in Table \ref{tab1} some properties of the investigated
reactions are summarized. 

The whole setup was housed inside the
Ciclope scattering chamber of INFN-LNS.  
The FAZIA blocks have been described in detail elsewhere (see for
example \cite{BougaultFAZIA,Valdre19,PastoreNIM,Pasquali2014,Carboni} and
references therein). Here we briefly recall that each block
includes 16 three-stage telescopes (300 $\mu$m-thick silicon  -  500
$\mu$m-thick silicon - 10 cm-thick CsI(Tl)).  
The preamplifiers and the fast digitizing electronic boards
operate under vacuum with proper cooling.  State of the art
capabilities in terms of isotopic identification have been achieved
for these detectors \cite{Carboni,PastoreNIM}.

Because of the strongly reversed kinematics of the
investigated reactions, the detection efficiency of the
setup is favoured at least for medium-heavy ejectiles. In fact, the efficiency for big fragments (Z$\geq$10)  is around
34\%, 41\% and 19\% for  $^{40}$Ca at 25 MeV/nucleon, $^{48}$Ca at 25 MeV/nucleon ~and
$^{48}$Ca at 40 MeV/nucleon, respectively; these values have been estimated by means of a simulation based on the
HIPSE code \cite{HIPSE}, filtered via a software replica of the setup
which takes 
into account the geometry and the identification
thresholds. 
The global efficiency for all the reaction channels 
is significant (around 41\% for $^{40}$Ca at 25 MeV/nucleon, around 46\% for
$^{48}$Ca at 25 MeV/nucleon and around 30\% for $^{48}$Ca at 40 MeV/nucleon). On the contrary, the efficiency for light particles is limited (around 2\% for Z=1 and around 4-5\% for Z=2 for all systems) as expected, due to their much broader angular distribution.
The global efficiency has been calculated as the fraction of simulated events surviving the experimental filter with at least one detected ejectile independently of the reaction mechanism, while that for big fragments (light particles) has been obtained as the fraction of simulated fragments with $Z\geq$10 (particles with Z=1,2) surviving the experimental filter.
As a consequence, although the detection of
complete events (in charge) is hindered and we cannot attempt
selections and analyses based on the event completeness,
 medium-heavy ejectiles can still be reliably characterized. 

\section{Data Analysis}
\subsection{A general overview of the reactions}
The studied systems are strongly mass asymmetric and thus 
- as it is shown in Table \ref{tab1} - the c.m. velocity is close to the projectile velocity.
 On the one hand this helps focussing most of the ejecta in a 
  narrow forward cone, thus enlarging the acceptance  in the c.m. frame,
 as quantified above. On the other
  hand the velocities of the heaviest  fragments, from peripheral
  collisions to more central ones, 
span a narrow interval.

\begin{figure*}[htpb]
\includegraphics[width=\textwidth]{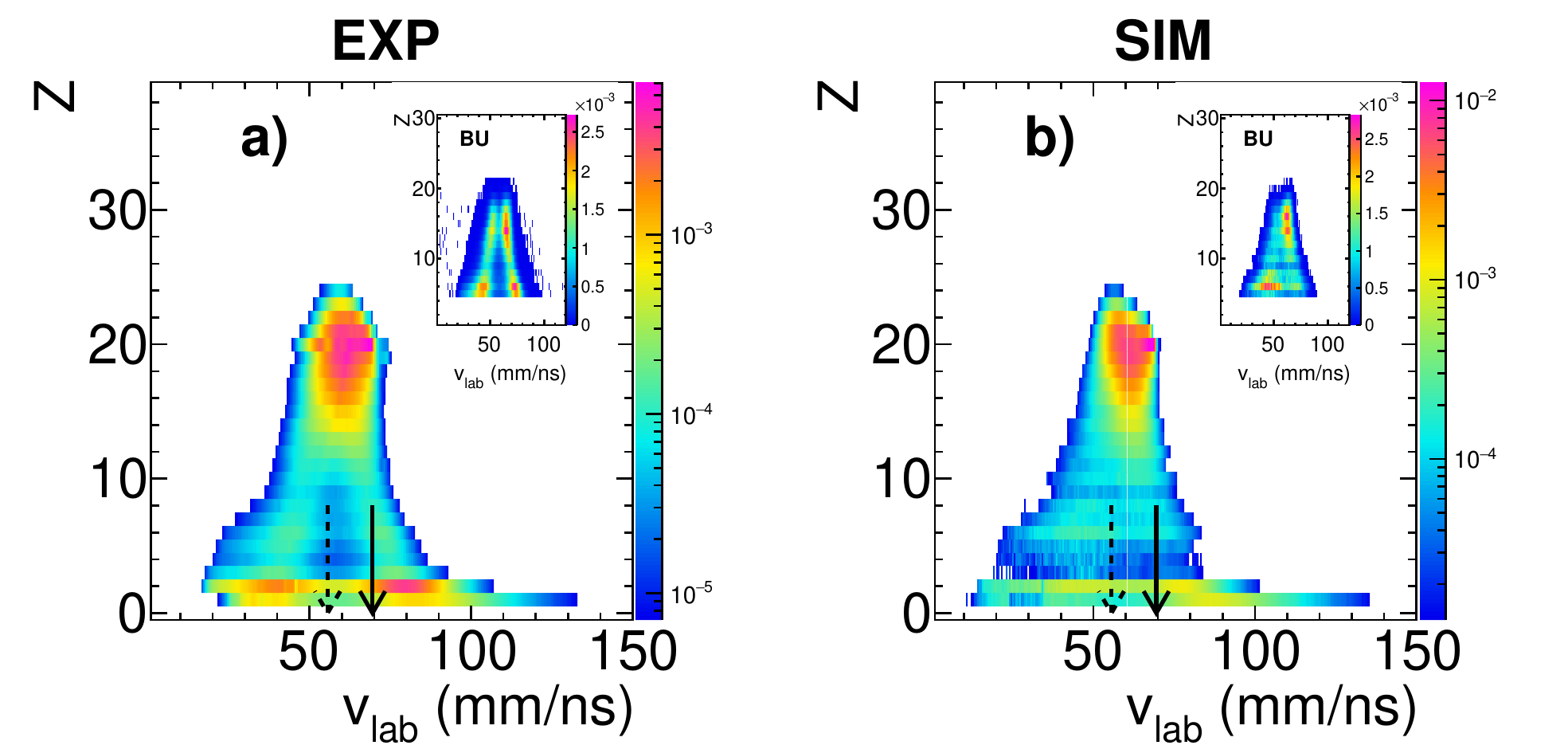}
\includegraphics[width=\textwidth]{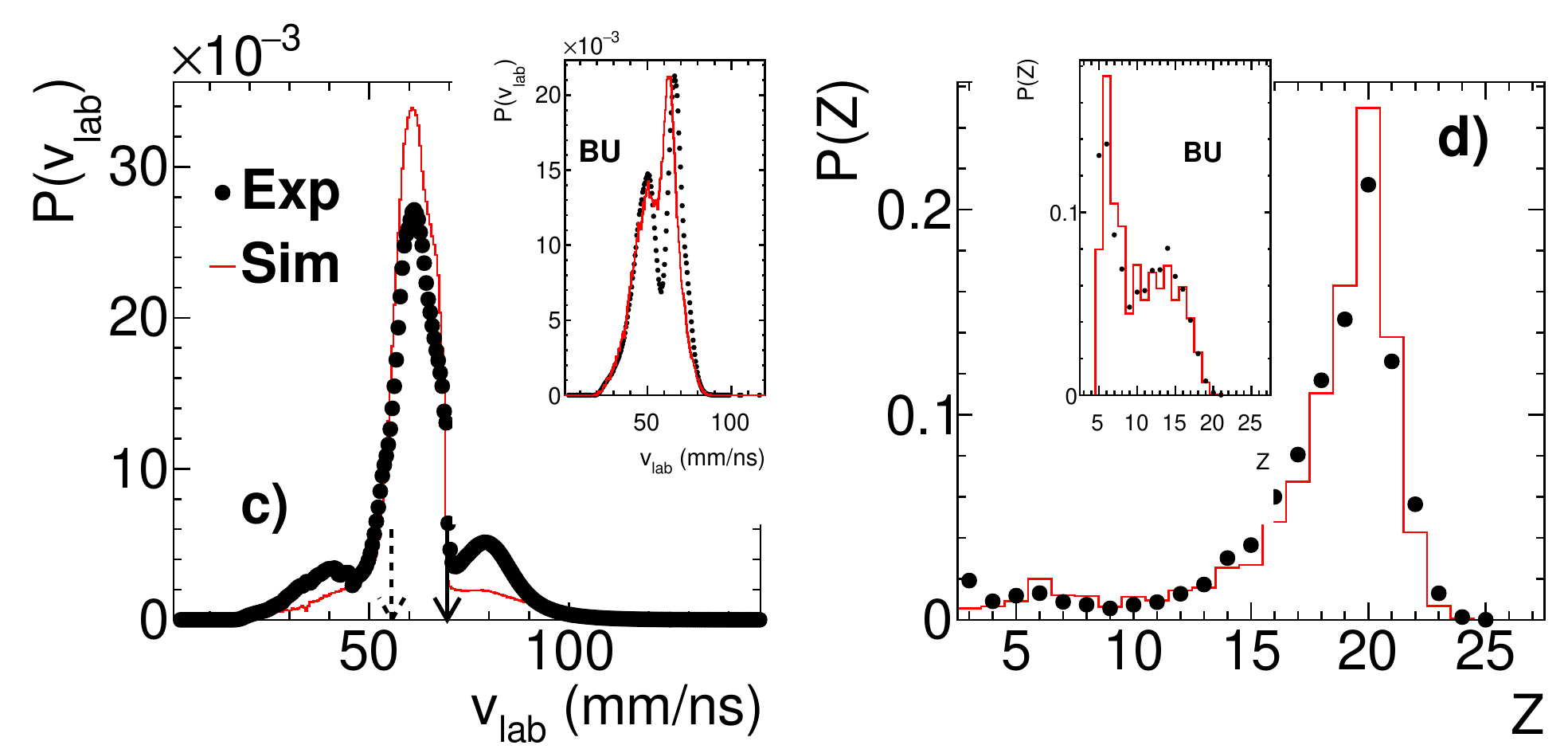}
\caption{$^{48}$Ca+$^{12}$C at 25 MeV/nucleon: (a) Charge-lab velocity
  correlation for all the ejectiles, experimental data; in the inset the correlation for the biggest and the second biggest fragment of the event for the BU selection is shown. (b) The same as in panel (a) but for  AMD+GEMINI++ simulation filtered via a software replica of the setup; (c) Relative probability distribution of lab velocity for all the ejectiles, symbols (red histogram): experimental
  (simulated) data. The dotted (continuous) arrow corresponds to the
  c.m. (beam) velocity. In the inset the projection on the velocity axis of the insets of panels (a) and (b) is shown.  (d) Relative probability distribution of charge for $Z>2$, symbols
  (red histogram): experimental (simulated) data. In the inset the projection on the charge axis of the insets of panels (a) and (b) is shown.} 
\label{fig1}
\end{figure*}

 This is illustrated in Fig.~\ref{fig1}, where, in panel (a),
the experimental charge-lab velocity correlation for all reaction products
is shown for the $^{48}$Ca at 25 MeV/nucleon~beam; the corresponding velocity and
charge distributions are shown in panel (c) and (d) (in panel (d) the charge distribution is presented for Z$>$2).
A spot located between c.m. (dotted arrow) and beam 
velocity (continuous arrow), with $Z\gtrsim15$ and  peaked around
$Z$=20 (as apparent in 
panel (d)) and with velocity almost
independent of the  charge, is clearly evident; this region mainly
corresponds to more or less damped events. The
quasi-elastic  
reactions producing weakly excited QT and QP nuclei
are not evident in these data, because their detection is significantly reduced by the angular coverage of the setup. In fact the
grazing angles (see Table \ref{tab1}) are well below the minimum angle covered by the array for all the reactions.
For lower
charges ($Z<14$) two branches emerge, one at
$v_{lab}>v_{c.m}$ and one at  $v_{lab}<v_{c.m}$, compatible with  the breakup (henceforth 
BU) of the heavier 
nuclei around Calcium produced by strongly damped collisions.
This is better
put into evidence in the inset of Fig.~\ref{fig1} panel (a), where the
charge-lab velocity correlation for the biggest and the second biggest
fragment of the event, both of them with $Z\geq 5$, according to
the BU selection introduced in subsection \ref{BU},
is shown. The corresponding lab velocity and charge distributions are
presented in the insets of panel (c) and (d). Two asymmetric branches,
with the one at velocities greater than $v_{c.m.}$ being more populated, are
apparent in the inset of panel (c). In the charge distribution (inset of
panel (d)) two regions emerge, basically corresponding to the light (below Z=10) and the heavy (beyond Z=10) BU fragment, 
as it will be better shown in
Fig.~\ref{fig3}. The efficiency for BU events, estimated as the fraction of simulated events with at least two fragments with $Z\geq5$ surviving the experimental filter, is below 1\% for all the reactions. 
The efficiency for the events in which a $Z\geq5$ fragment is accompanied only by light ejectiles with $Z<5$ is around 8-10\% for all reactions and it decreases by about 1\% when requiring that the charge of the heavy residue is $Z\geq$10 and that it is isotopically identified.

In order to shed more light on the reaction mechanism, the
experimental data have been compared with a simulation based on the
the AMD code \cite{Ono92,Ono99,OnoJPC2013}
coupled to GEMINI++ \cite{Charity10} as an afterburner, as already
successfully done to study other FAZIA experiments ~\cite{Piantelli2020,Piantelli2021,Camaiani2021,Frosin2022,Ciampi2022}.  Details
concerning the used version of the code can be found in
\cite{FIASCO19,Piantelli2020}. 
 The simulated data, filtered through a
software replica of the setup, are shown in
Fig.~\ref{fig1} (b), (c) (red histogram) and (d) (red histogram) and reasonably
reproduce the experimental
ones. In fact the same spot at high Z as in the experimental case is present and 
the overall charge distribution is well reproduced (as shown in panel (d) for Z$>$2). However, the
two velocity branches, especially in the low-Z region from Lithium to Boron,
are less separated, as apparent in the inset of panel (c) when the BU selection is applied. In particular, the branch at velocity greater than $v_{c.m.}$ is shifted towards lower values with respect to the experimental case. On the other side, the charge distribution is reasonably reproduced also for BU events (inset of panel (d)).  

\begin{figure}[htpb]
\includegraphics[width=0.3\textwidth]{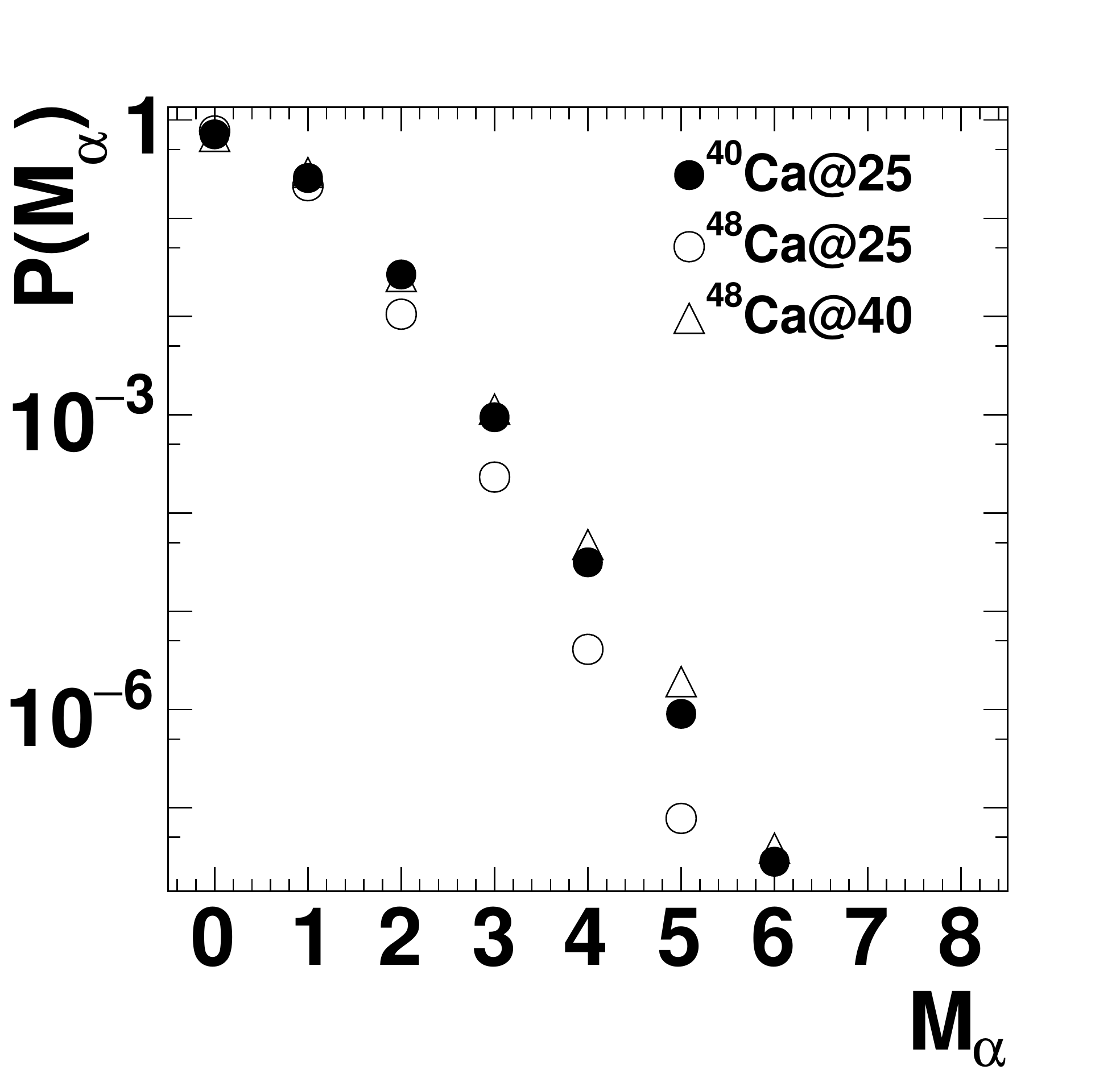}
\caption{Relative probability distribution of $\alpha$ particle multiplicity
  for all the measured reactions.}
\label{fig2}
\end{figure}

Concerning light products, channels involving $\alpha$ particles are expected to be 
abundant in reactions
with a $^{12}$C target in particular for the
N=Z projectile case~\cite{raduta_plb_2011,borderie_plb_2016}. In fact, as shown in Fig.~\ref{fig2},  
events with at least 3 $\alpha$ represent about $0.1\%$ of the total, thus they are sufficient to study $\alpha$ 
particle  correlations,  such as - for
example - those coming from the de-excitation of $^{12}$C$^{*}$. The picture shows that, although the detection efficiency for $\alpha$ particles are very similar for all the reactions, in case of
  $^{48}$Ca beams events with many $\alpha$ are more
  abundant for the reaction at 40 MeV/nucleon than for that at 25
  MeV/nucleon, due to the higher available energy favouring in
  general nuclear fragmentation. The $^{40}$Ca beam, although at 25 MeV/nucleon, shows
  $\alpha$ multiplicities close to the $^{48}$Ca at 40 MeV/nucleon case, probably due to the nature of the system where both target and projectile are $\alpha$-conjugate. 

\subsection{Breakup (BU) channel}
\label{BU}

Keeping in mind that the collected events are mainly related to
strongly damped collisions with the formation of a heavy excited
source, the BU channel is
selected by requiring events with 
two fragments with Z$\geq$5 each, labelled as $HF$ (heavy fragment, with charge Z$_{HF}$ and mass A$_{HF}$)
and $LF$ (light fragment, with charge Z$_{LF}$ and mass A$_{LF}$). The associated reconstructed parent nucleus (henceforth denoted as $BU$ fragment or $BUF$, with charge Z$_{BU}$=Z$_{HF}$+Z$_{LF}$ and mass A$_{BU}$=A$_{HF}$+A$_{LF}$) has
therefore Z$\geq$10, a value 
roughly corresponding to the lower limit of the peak of the charge distribution
of the biggest detected fragments (see Fig. \ref{fig1} (d)). 

Before discussing in detail the properties of the BU channel, we
mention that in the experimental sample
the fraction of BU events with respect to
the number of events  (from now on labelled as ``residue'' or R events) with one fragment with Z$\ge$10 (henceforth called residue or R fragment, with charge Z$_R$ and mass A$_R$) and at least another ejectile with Z$<$5 is 14-15\% for all reactions. 

\subsubsection{Source size and asymmetry of the split}

\begin{figure}[htpb]
\begin{tabular}{cc}
\includegraphics[width=0.25\textwidth]{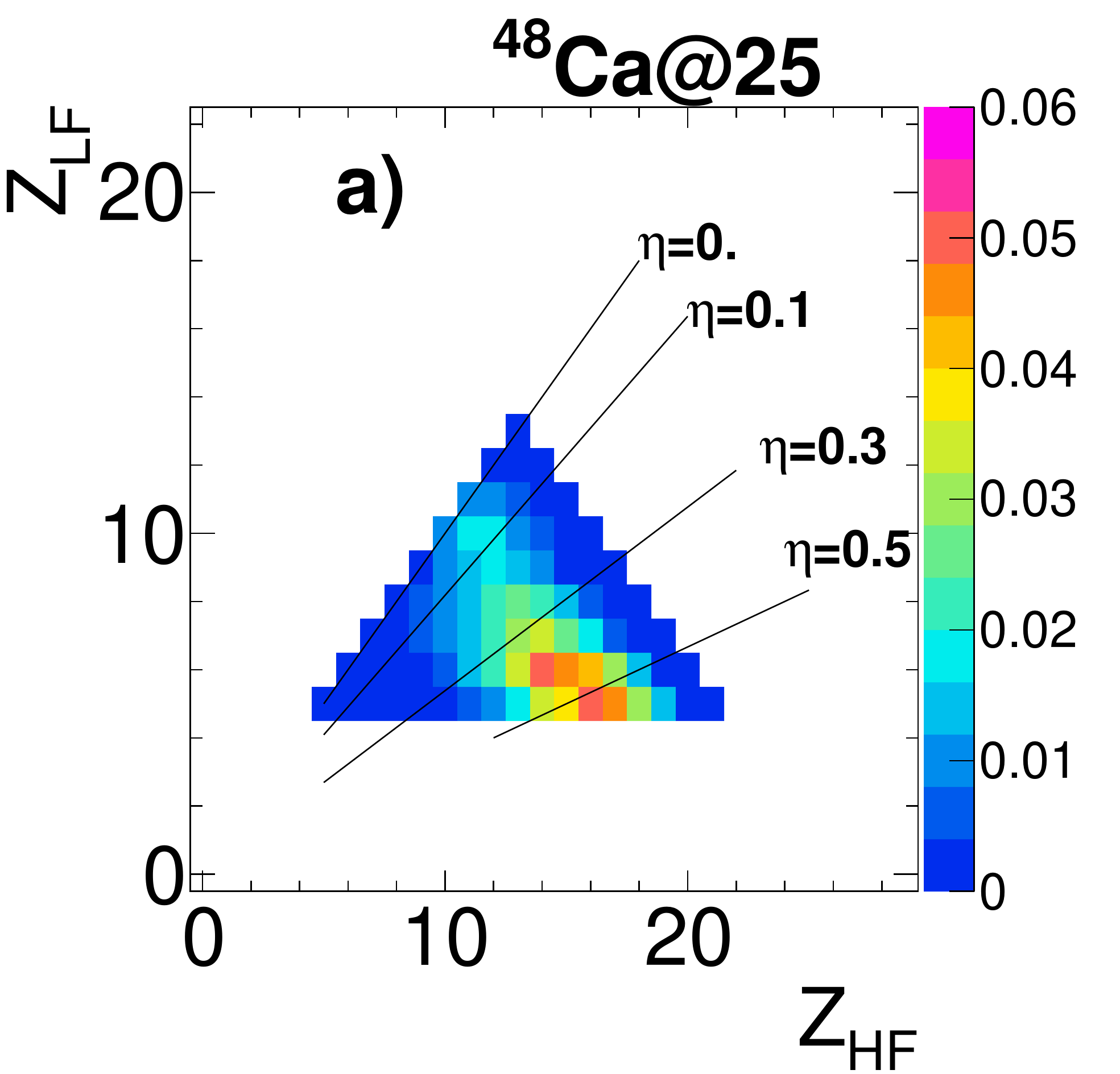}&\includegraphics[width=0.25\textwidth]{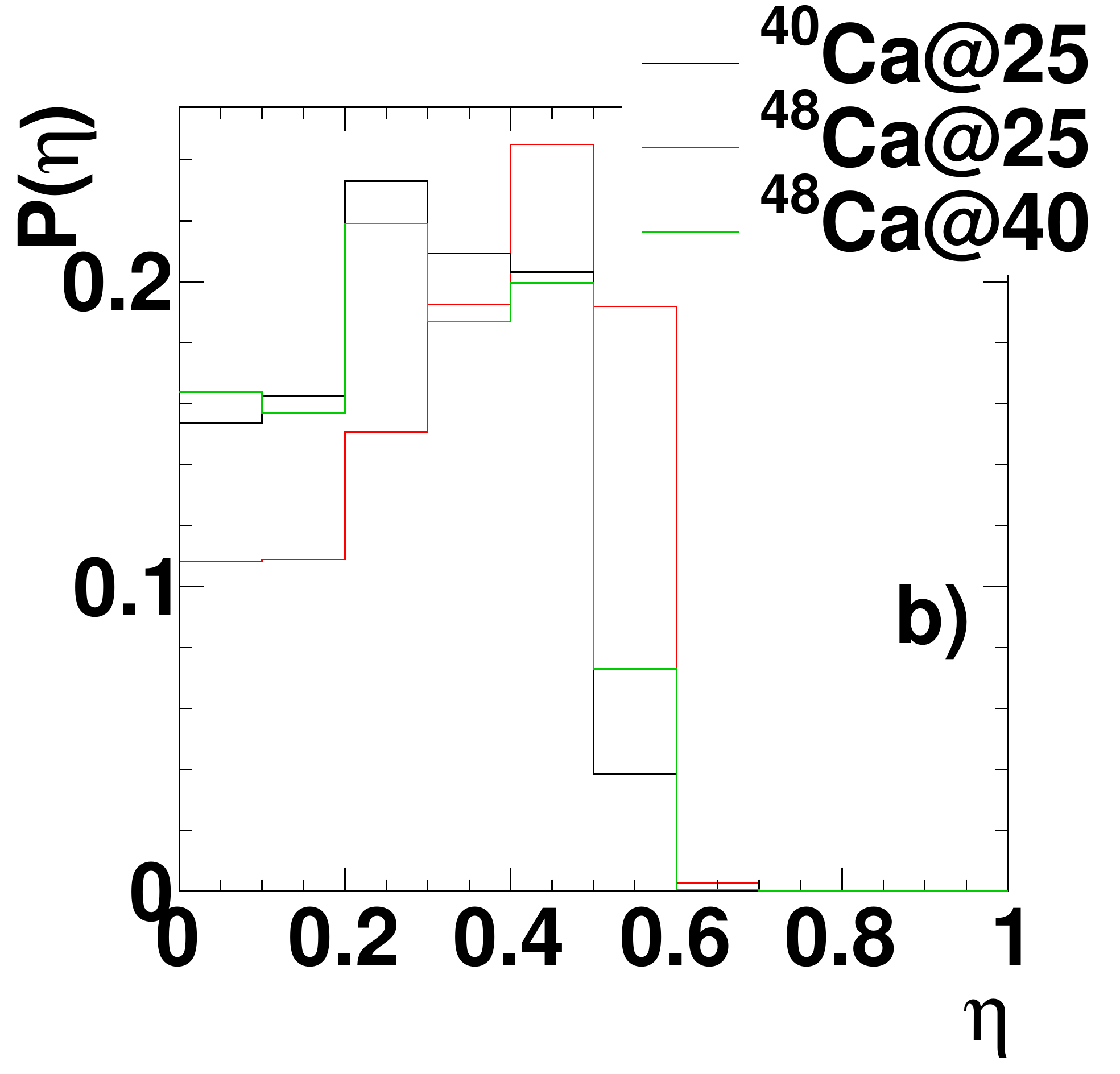}\\
\includegraphics[width=0.25\textwidth]{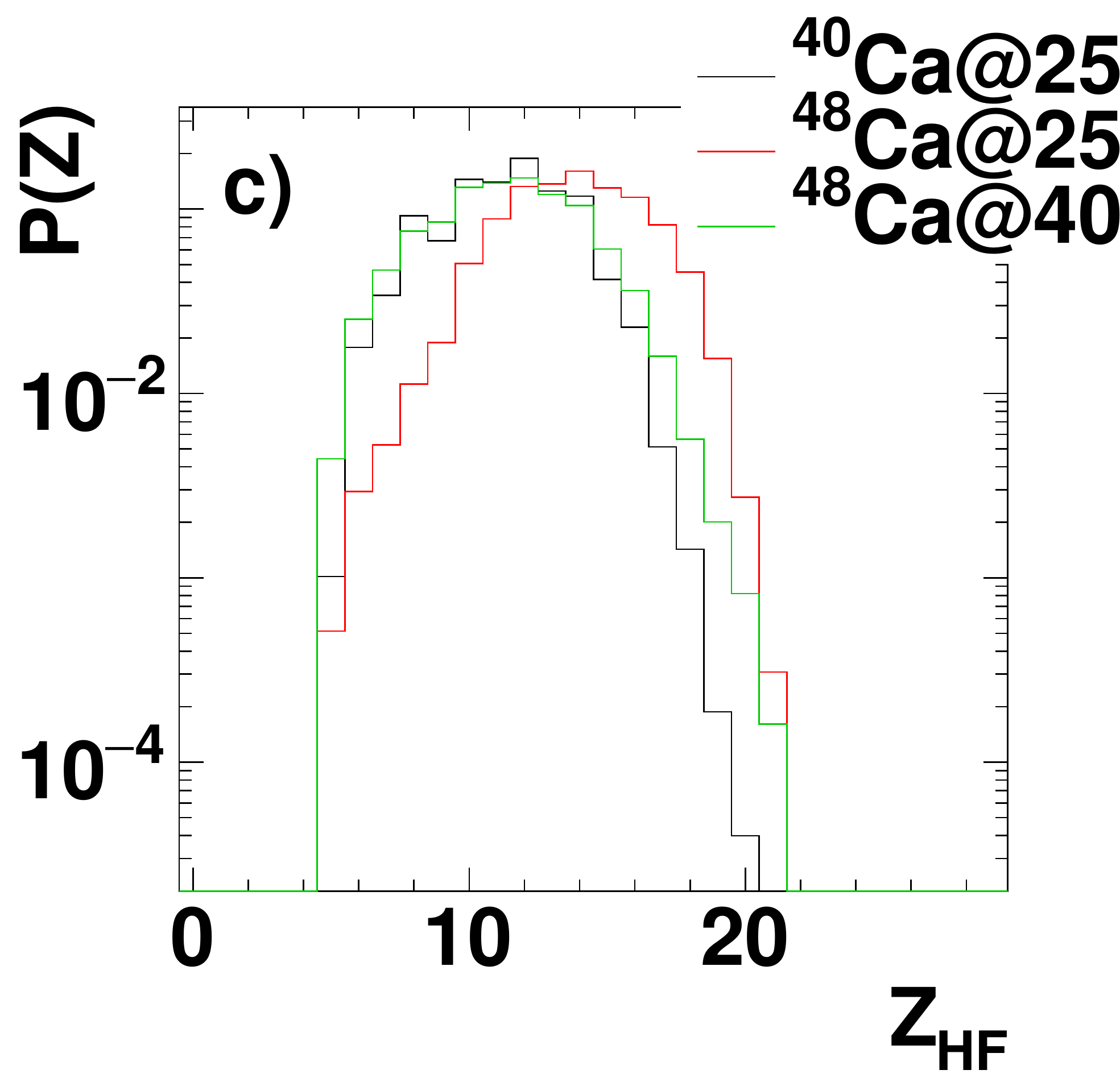}&\includegraphics[width=0.25\textwidth]{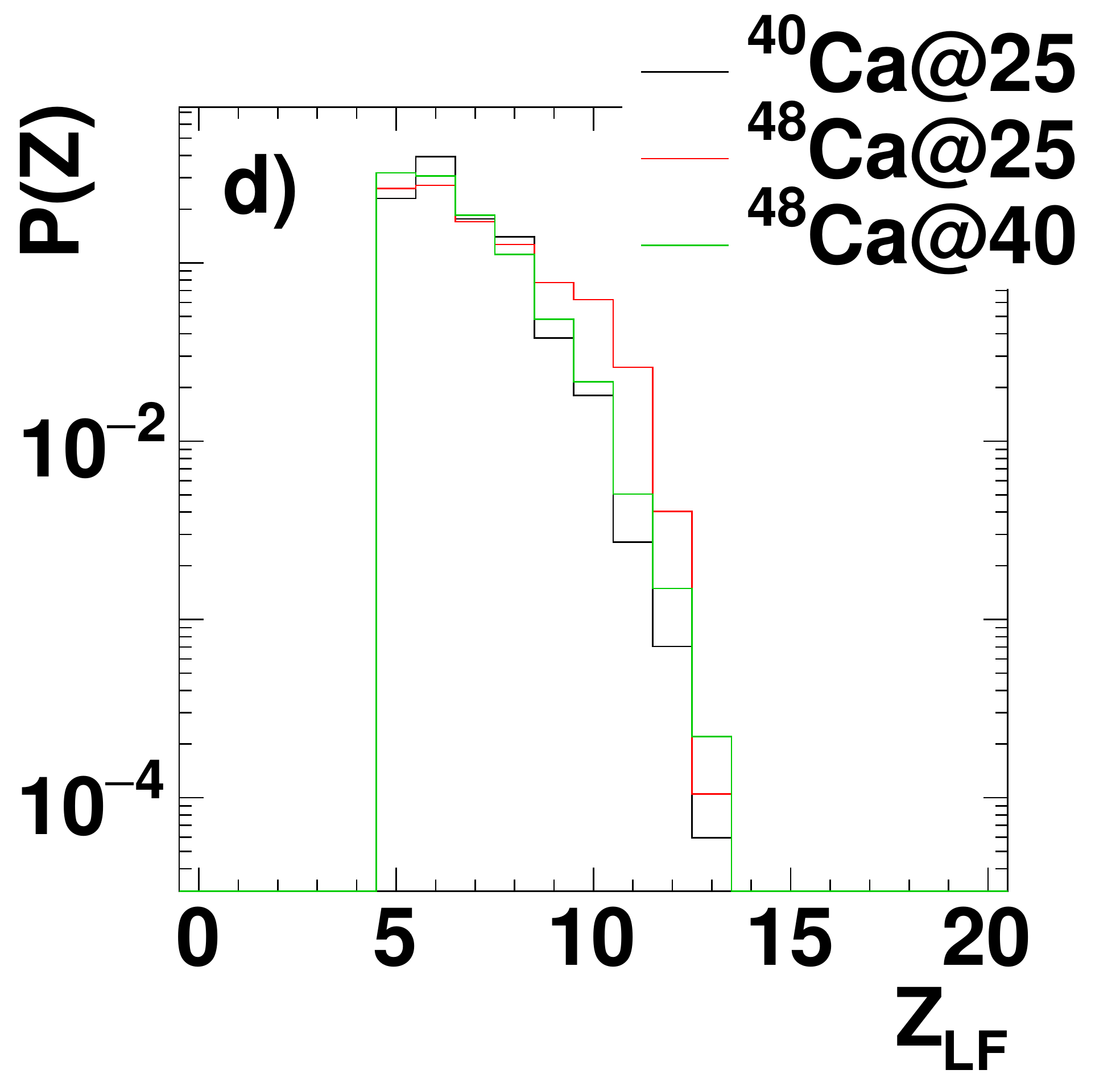}\\
\end{tabular}
\caption{a): Z$_{LF}$ vs. Z$_{HF}$ correlation for $^{48}$Ca at 25
  MeV/nucleon. For all systems, relative probability distribution of : b) $\eta$; c) Z$_{HF}$;
   d) Z$_{LF}$.} 
\label{fig3}
\end{figure}

For all systems the BU favours asymmetric splits as expected for the
binary division of light nuclei (fission-like channel, where surface and Coulomb
energies dominate the process), quite below
the Businaro-Gallone region. An example
of this is shown in Fig. \ref{fig3} (a), which presents the Z$_{LF}$ vs Z$_{HF}$ plot for the $^{48}$Ca+$^{12}$C
reaction at the lower energy;  
definining $\eta$ as the charge
asymmetry  (i.e. \(\eta=\frac{\mathrm{Z}_{HF}-\mathrm{Z}_{LF}}{\mathrm{Z}_{HF}+\mathrm{Z}_{LF}}\)),
the BU yield concentrates at $\eta>0.3$. More
quantitatively Fig. \ref{fig3} (b)   compares the
experimental asymmetries for the BU measured in the
three reactions.  The most asymmetric BU occurs for the neutron rich
system at the lower energy while at 40 MeV/nucleon~the  $^{48}$Ca  reaction produces
an asymmetry  comparable 
with that of  $^{40}$Ca beam collisions at low energy. The Fig. \ref{fig3} (c) and
(d) separately show the Z$_{HF}$ and Z$_{LF}$ distributions, for the three
systems. A hierarchy can be seen: mainly for $HF$, but also to a lesser extent for $LF$, the distributions obtained for the $^{48}$Ca beam   
at 25 MeV/nucleon are shifted towards higher values with respect to the other cases. The $^{48}$Ca beam at 40 MeV/nucleon gives values slightly heavier than  the $^{40}$Ca case. These effects are possibly due to the fact that the $^{48}$Ca beam, being more neutron rich, can keep bound larger fragments with respect to the $^{40}$Ca case, but when the energy increases longer decay chains are expected due to the higher excitation energy deposited in the system; also the dynamical emissions of particles (most likely, but not exclusively, neutrons for the n-rich projectile), again increasing with the beam energy, contribute to decrease the final fragment charge.

\begin{figure*}[ht]
\begin{tabular}{cc}
\includegraphics[width=0.5\textwidth]{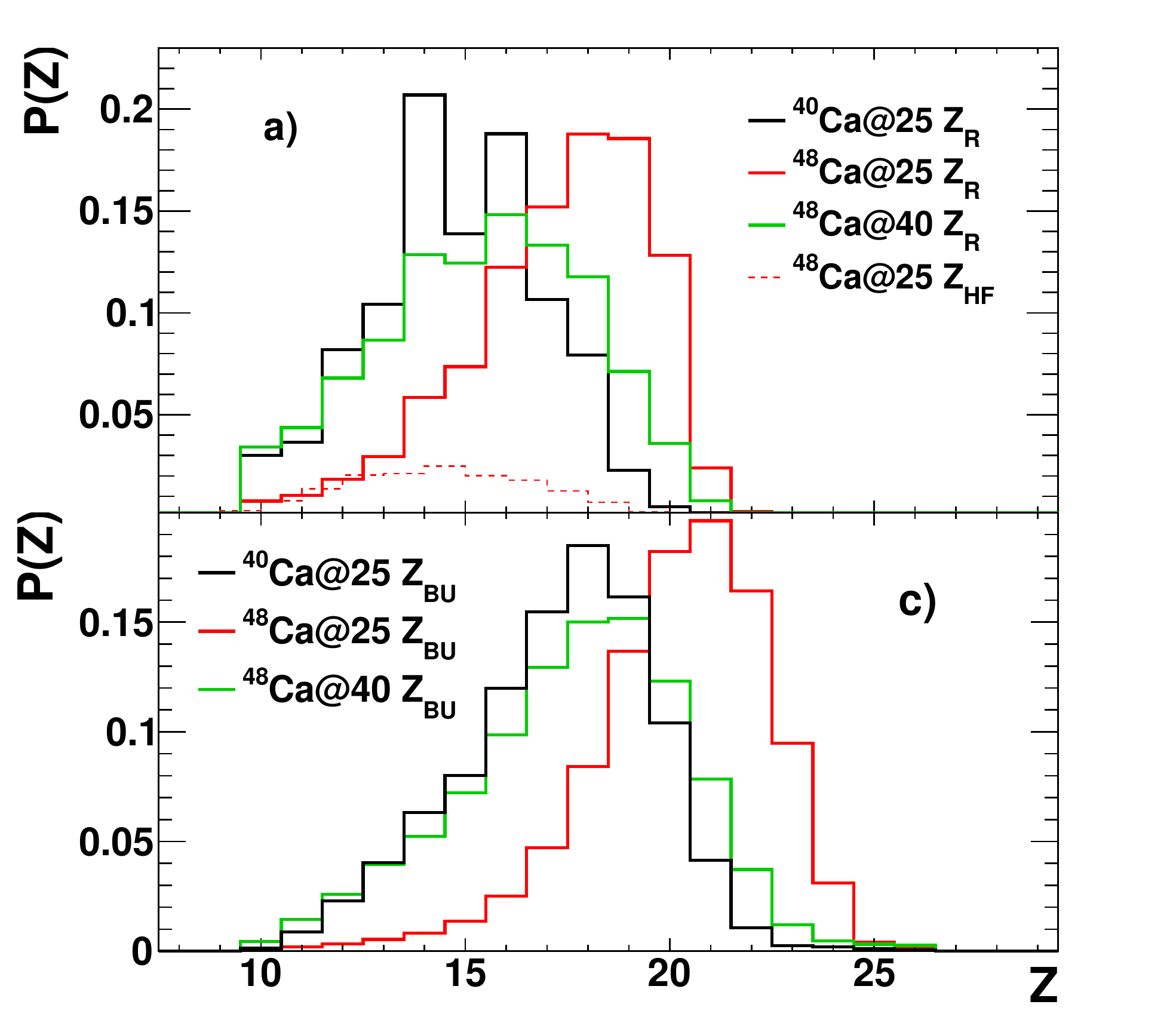}&\includegraphics[width=0.5\textwidth]{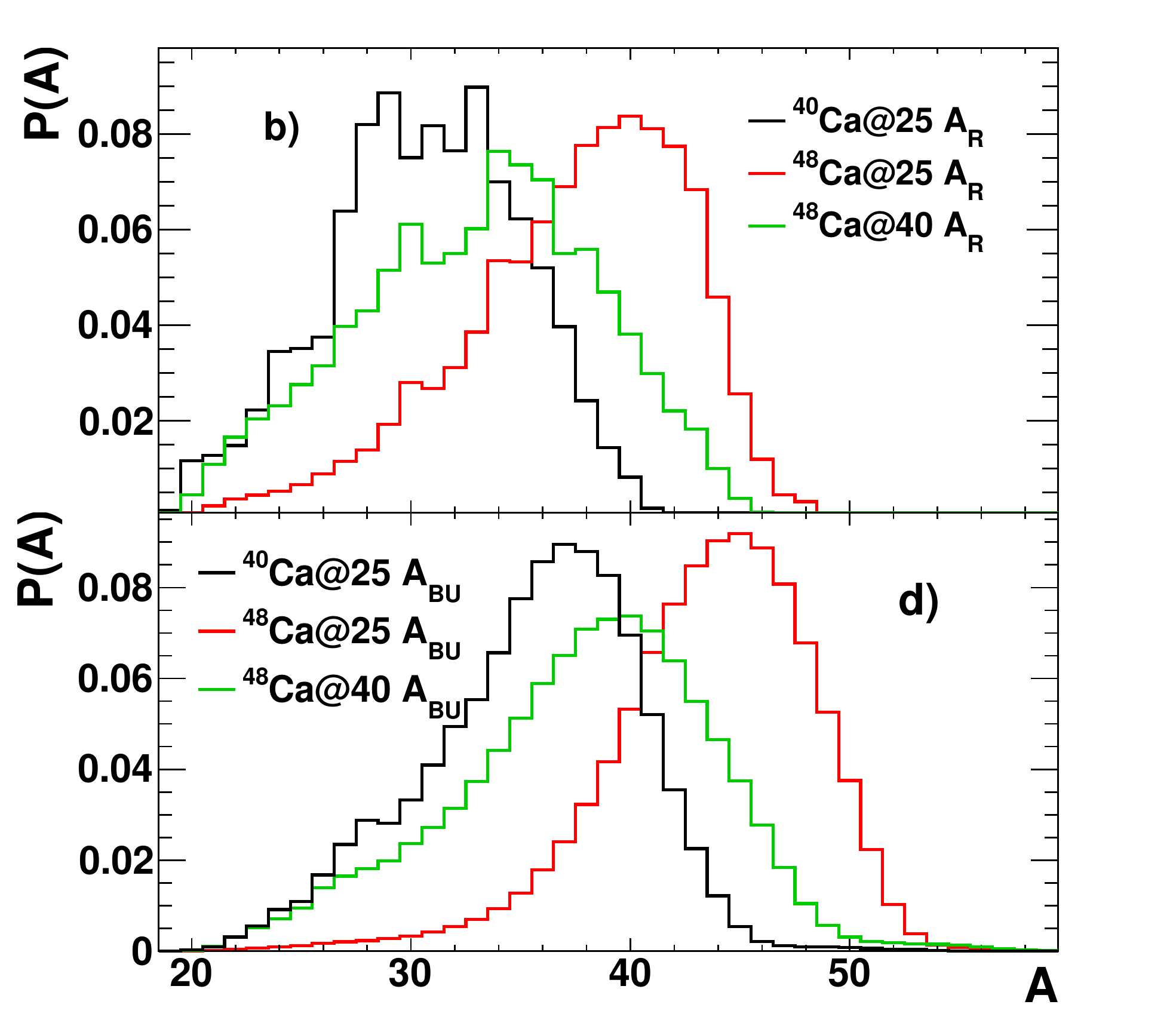}\\
\end{tabular}
\caption{a): Relative probability distribution of Z$_{R}$ for all the systems. Only the subset of events for which the mass of the heavy ejectile is detected is shown. Dashed histogram: Z$_{HF}$ distribution. b) Relative probability distribution of mass (A$_{R}$) for the same fragments of panel (a). c) Relative probability distribution of Z$_{BU}$. Only the subset of events for which the mass is experimentally detected for both
  fission fragments is shown. d) Relative probability distribution of mass (A$_{BU}$) for the same events of panel (c).}
\label{fcm}
\end{figure*}

Similar considerations can be applied also to the Z$_{R}$ (and A$_R$) distributions for residue events 
shown in  Fig. \ref{fcm} panels (a) and (b), where the same hierarchy found for Z$_{HF}$ and Z$_{LF}$ (and also for Z$_{BU}$ and A$_{BU}$, panels (c) and (d) of  Fig. \ref{fcm}) is observed; in this figure only events in which all the involved fragments are identified both in charge and in mass are included; for the BU channel this corresponds to 63\%, 82\% and 69\% of all the BU events for the three reactions, respectively, while for the R channel this corresponds to 73\% for the $^{40}$Ca case, and to 88\% for both the $^{48}$Ca beams, of all the residue events. The average values obtained for Z$_{BU}$, A$_{BU}$, Z$_{R}$ and A$_R$ are reported in Table \ref{tabaz}. Both the average values of the table and the distributions of Fig. \ref{fcm} show that the reconstructed fragments in the BU case are heavier than the residue ones for a given reaction. This may indicate that in case of BU part of the available excitation energy is exploited for the breakup process\footnote{on the basis of the obtained fragment distributions, we can estimate that typical $Q_{values}$ for the splitting are of the order of -25 MeV}, thus reducing the deexcitation chain or that breakup process selects by itself heavier primary fragments.  Due to the incomplete angular coverage of the setup, the Z$_{R}$ distribution may include also $HF$'s for which the $LF$ has been lost; however, such a contribution, unavoidably present, is limited to the low Z tail of the Z$_{R}$ distribution, as shown in panel (a) by the red dashed line (corresponding to the Z$_{HF}$ distribution for $^{48}$Ca at 25 MeV/nucleon). In Table \ref{tabaz} also the average values of the charge and mass distributions obtained for simulated data are presented, showing that the experimental hierarchy is reasonably reproduced by the simulation, with all the predicted values very close to the experimental ones.

\begin{table*}[tb]
\begin{tabular}{|c|c|c|c|c|c|c|c|c|c|}
\hline
Projectile & Energy & \multicolumn{2}{c|}{average Z$_{BU}$} & \multicolumn{2}{c|}{average Z$_R$} &  \multicolumn{2}{c|}{average A$_{BU}$} & \multicolumn{2}{c|}{average A$_R$} \\
&MeV/nucleon&EXP&SIM&EXP&SIM&EXP&SIM&EXP&SIM\\
\hline
$^{40}$Ca&25&17.2$\pm$0.4 &16.8$\pm$0.4&14.8$\pm$0.4 &15.0$\pm$0.5&35.5$\pm$0.9 &34.4$\pm$0.9&30.6$\pm$0.9 &31.0$\pm$1.0\\
\hline
$^{48}$Ca&25&20.2$\pm$0.4 &19.8$\pm$0.4&17.3$\pm$0.4 &17.0$\pm$0.4&43.6$\pm$0.8 &42.5$\pm$0.9&37.5$\pm$0.9& 36.5$\pm$0.9\\
\hline
$^{48}$Ca&40&17.7$\pm$0.4& 17.0$\pm$0.4&15.4$\pm$0.4 &15.7$\pm$0.5&37.9$\pm$0.9 &36.5$\pm$0.8&33.0$\pm$0.9 &33.5$\pm$0.9\\
\hline
\end{tabular}
\caption{Average charge and mass values for the measured heavy fragment in residue events (R) and for the reconstructed nucleus (obtained adding Z$_{LF}$ and Z$_{HF}$ for the charge and A$_{LF}$ and A$_{HF}$ for the mass) in BU events (BU), both for experimental and simulated data.}
\label{tabaz}
\end{table*} 

The obtained
results for the reactions at 25 MeV/nucleon~concerning the scaling
with the neutron richness of the system of both
the BU asymmetry  and the heaviest fragment size are fully compatible with the results on Ca+Ca,Ti reactions at the
same energy reported in~\cite{amorini_prl102_2009}, although not exactly comparable: indeed, the neutron
richness of the system allows for the formation 
of heavier
fragments that, in case of BU, produce more asymmetric splits.\\

\begin{figure}[htpb]
\includegraphics[width=0.5\textwidth]{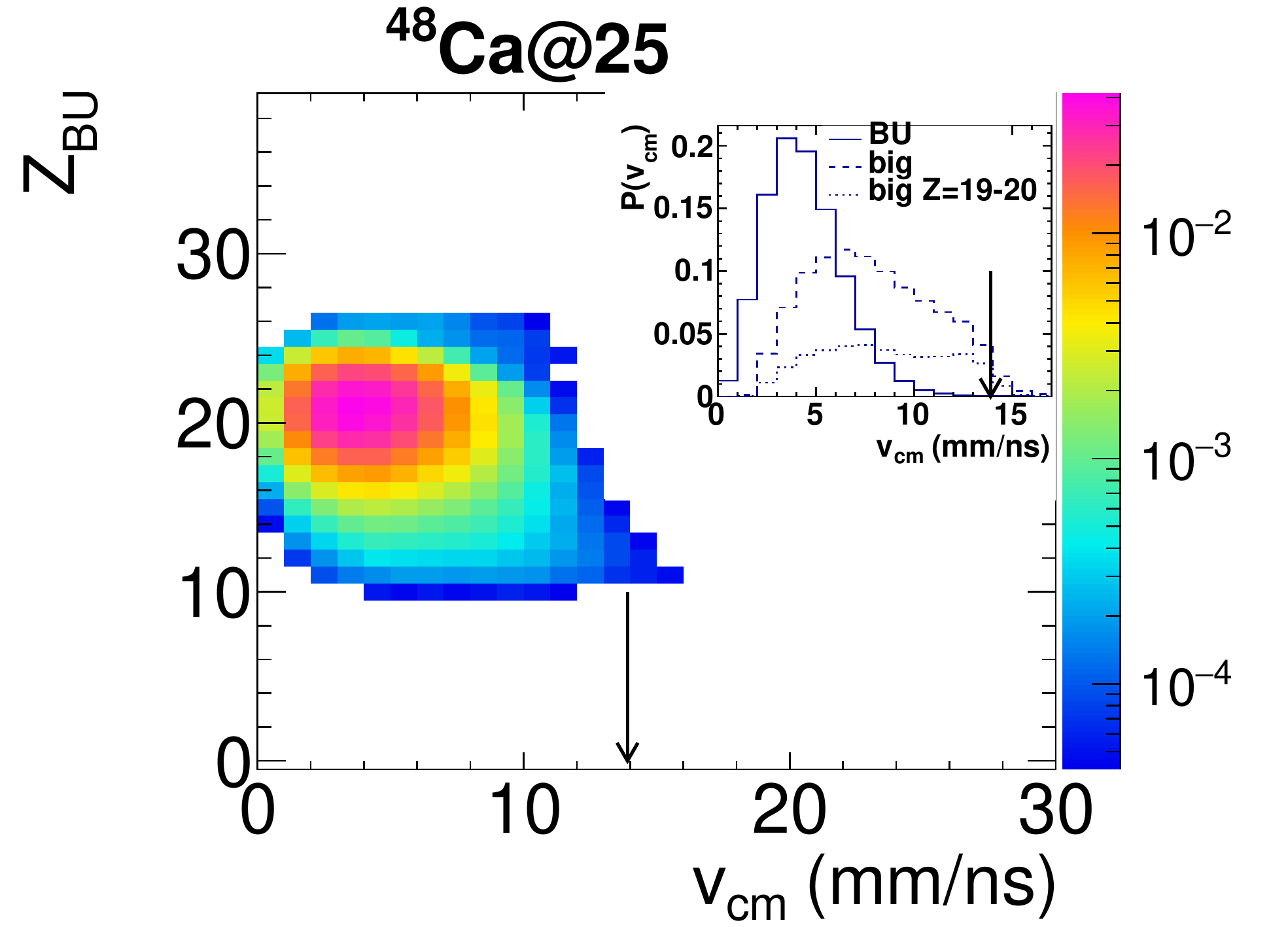}
\caption{$Z_{BU}$ vs. $v_{cm}$ plot (reconstructed from the two
  breakup fragments) for the system $^{48}$Ca at 25 MeV/nucleon. The
  arrow corresponds to the beam velocity. In the inset the projection on
  the $v_{cm}$ axis (continuous histogram) is shown, together with the
  c.m. velocity distribution of the biggest fragment of the event with $Z\geq10$
  (dashed histogram) and with $Z=19-20$ (dotted histogram);
  distributions normalized to their integral. 
}  
\label{fig5emezzo}
\end{figure}

In Fig. \ref{fig5emezzo} we show Z$_{BU}$ vs. the reconstructed c.m. velocity of the BU fragment\footnote{i.e. the c.m. velocity of the centre of mass of $HF$ and $LF$}
for
the reaction $^{48}$Ca+$^{12}$C at 25 MeV/nucleon{}, when both fragments are isotopically identified: the
velocity of the reconstructed nucleus lies
in between that of the c.m. and that of the
projectile (arrow position), closer to the former,  signaling a
strong degree of dissipation.  This is  similar
to what can be recognized in Fig. \ref{fig1} (a), where heavy
residues have lab velocities  between the c.m. and the beam velocity; to
better illustrate this point, 
in the inset of  Fig. \ref{fig5emezzo} we show the
c.m. velocity distribution of all the BUFs as
continuous histogram, together with the c.m. velocity of the biggest
fragment of the event, both for the class  Z$\geq$10 (dashed
histogram) and for the specific channel
Z=19-20 (dotted histogram).
The distribution of the detected biggest fragment with Z$\geq$10 
is rather broad and slightly shifted to velocities higher than that obtained for the
BUF, thus signaling that a larger degree of
dissipation is associated with the BU process. Moreover, as expected, when the
charge of the biggest fragment is close to the projectile one, there is
also a contribution in the region of the projectile velocity,
associated with peripheral reactions,  lacking at all in the BU case.
Therefore, it is quite reasonable to conclude that the
candidate BU fragments 
 are indeed compatible with a breakup process occurring after a 
very dissipative collision producing an incomplete fused system or an
excited QP. 
In this respect we mention that the
  systematics~\cite{eud14} predicts for our systems
a relative abundance of 15-17\% at 25 MeV/nucleon
and around 8\% at 40 MeV/nucleon for the fusion channel.

\subsubsection{Isotopic composition of the BU fragments}

\begin{figure}[htpb]
\includegraphics[width=0.4\textwidth]{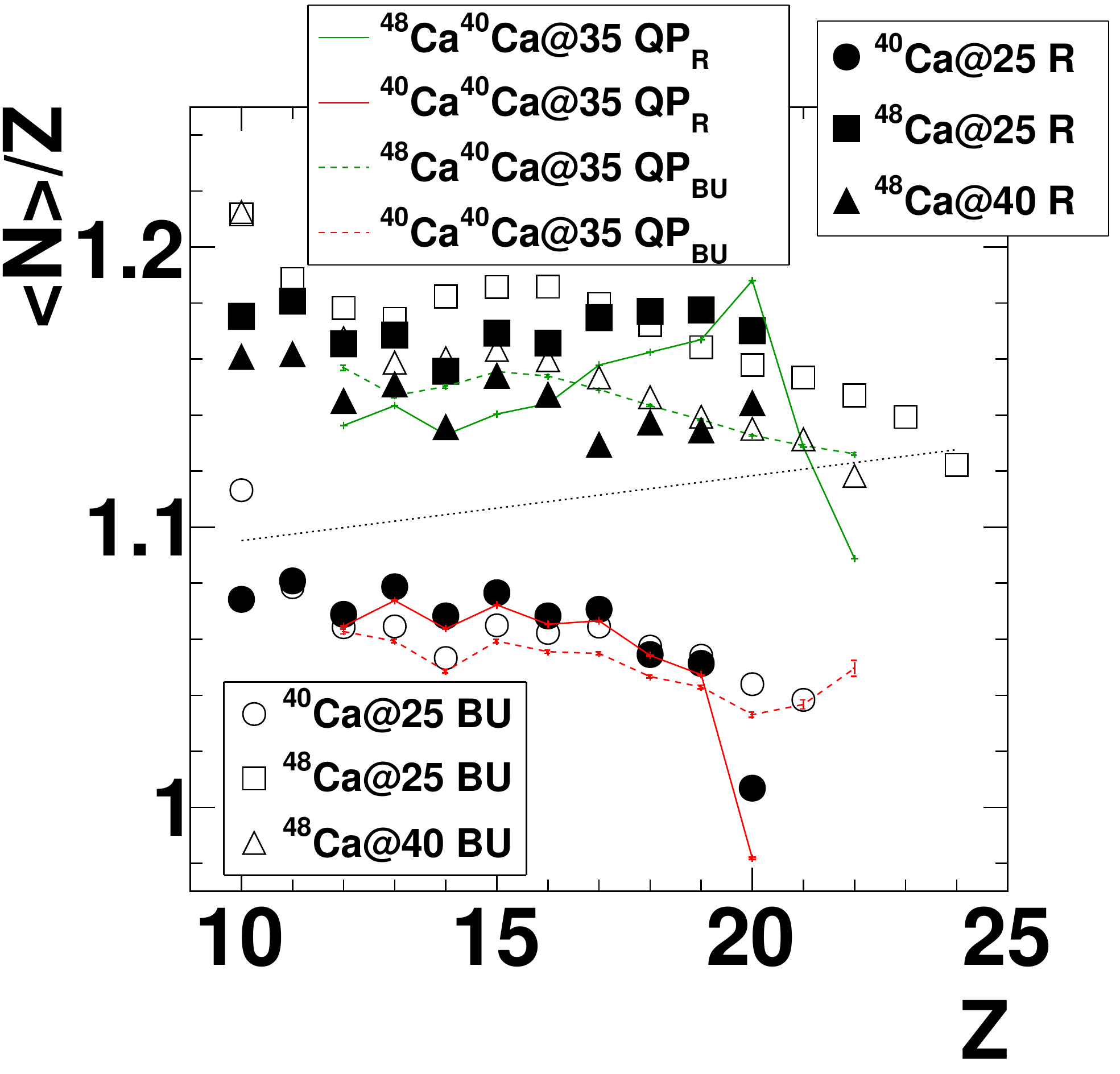}
\caption{Average N/Z as a function of Z; open symbols reconstructed BU fragment, full symbols R fragment. Dashed and continuous lines refer to reconstructed QP$_{BU}$ and QP$_R$, respectively,  from the data set of \cite{Camaiani2021}. 
  The dotted black line represents the EAL prediction (average
  N/Z 
  after statistical evaporation). Statistical errors are within the symbol size.} 
\label{fig4}
\end{figure}

We now discuss the isotopic content of the BUFs,
also in
comparison with that of the R fragment.
The average N/Z values  are shown in
Fig. \ref{fig4} as a function of the charge 
for the three reactions.

Two main observations can be given.
The first one, concerning the neutron content of the beams, is the expected  big gap separating the
neutron rich reactions from the N=Z one; the N/Z gap is around 0.1
corresponding at Z=20 to a difference of two neutrons. The n-rich
cases are essentially above the N/Z corresponding to the Evaporation Attractor Line (EAL)\cite{Charity98} (dotted line in
the figure) while the  $^{40}$Ca one is below that line. Assuming that the EAL line cannot be crossed during the statistical de-excitation phase, this could be interpreted as a sign that, independently of the dynamics of the reaction, the primary fragment population tends to remain on the same side of the EAL as the entrance channel.

A second comment concerns the beam energy: for the $^{48}$Ca beam,
 the average N/Z is larger for the 
25 MeV/nucleon reaction than for the 40 MeV/nucleon one. 

Considering the two channels (i.e. BU -open symbols- or single heavy residue -full symbols-) the differences are small. However, the  BU 
fragments are
slightly more neutron rich for Z$<$18 
for  the $^{48}$Ca projectile and slightly less neutron rich for the
$^{40}$Ca projectile with respect to the R fragments. 
The region beyond the projectile is not
populated in the R case, while it is accessed by summing $HF$ and $LF$ in the BU channel.  

The previous observations can be qualitatively understood by 
taking into account that, for primary sources with similar initial
values of N and Z, larger excitations and thus longer evaporation
chains lead closer to the EAL, which is approached by opposite injection points in
the N-Z map for the two Ca beams. 
Thus, also the slight N/Z differences shown in the two channels (BU and R)
might signal, on average, shorter evaporation chains for the
BU channel due to the energy consumed by the splitting process.
The net average N/Z decrease  (and the consequent closer approach to the EAL) when the beam energy increases for the $^{48}$Ca reactions can
again reflect the average larger excitation energies at 40 MeV/nucleon (triangles vs. squares) but 
an additional contribution from
pre-equilibrium (neutron richer) emissions cannot be ruled out.

Figure  \ref{fig4} also contains the average N/Z values of
quasi-projectile residues detected as such (QP$_R$, continuous lines) or reconstructed after breakup adding charge and mass of the two scission fragments (QP$_{BU}$, dashed lines),
measured in 
$^{48}$Ca+$^{40}$Ca and $^{40}$Ca+$^{40}$Ca collisions at
35 MeV/nucleon 
coming from the data set
investigated in \cite{Camaiani2021}. It is important to note that since the Ca+Ca case corresponds to symmetric or near symmetric reactions, it is easy to select the QP source, at variance with what happens for the Ca+C case. 
Therefore we can now compare the average N/Z obtained for the QP of the Ca+Ca systems with the results for Ca+C, keeping in mind that in the latter case we are dealing with very damped events, where QP and incomplete fusion are not easily distinguishable.

For $^{40}$Ca beams, where the N/Z of whole system and of the reaction partners separately is 1, the $\langle$N$\rangle$/Z of the symmetric and asymmetric reactions are similar both for the BU/QP$_{BU}$ channel (dashed red line and open circles) and for the R/QP$_{R}$ one (continuous red line and full circles). 

For $^{48}$Ca beams the situation is less clear, because in this case the N/Z values of the whole system are different for the two datasets (1.2 for $^{48}$Ca+$^{12}$C and 1.3 for $^{48}$Ca+$^{40}$Ca). It is evident that the Ca+Ca results (green lines) at 35 MeV/nucleon are closer to the Ca+C reaction at 40 MeV/nucleon (triangles), as reasonably expected. In the R channel the effect of peripheral collisions in the Ca+Ca case (continuous green line) pushes up the average N/Z towards the projectile value for Z close to 20, at variance with what happens for the Ca+C (full triangles), where such reactions are suppressed by the setup geometry. Instead  far from the peripheral region, for the QP$_{R}$/R channel the $\langle N\rangle /Z$ is almost independent of the target ($^{40}$Ca or $^{12}$C). A possible explanation of this observation may be related to the isospin diffusion for the $^{48}$Ca+$^{40}$Ca case, which tends to shift the average N/Z of the QP towards the value of the whole $^{48}$Ca+$^{12}$C system.

\subsubsection{The relative velocities of BU
pairs}

We now move on to investigate an important
observable of the BU process, namely
the relative velocity of the split partners. As a baseline we expect  a phenomenon mainly ruled by the Coulomb repulsion, where the asymptotic relative
velocity should be described by the Viola systematics for 
fission~\cite{Viola85,Jing99,Fan2000}, although dynamical effects may give additional contributions \cite{DeFilippo2012,defilpaga2014}.
For the following discussion we introduce upper and lower limits on Z$_{BU}$, in order to exclude on the one hand very small reconstructed parent nuclei missing a large part of the available charge and on the other the small tail of very big reconstructed fragments, corresponding to spurious events.
Therefore we adopt Z$_{BU}>$13 (corresponding to more than 50\% of the total charge of the entrance channel) as lower limit for all the systems; on the basis of the distributions of Fig. \ref{fcm} (c), we choose as upper
limits  Z$_{BU}$=24,22 for the reactions with $^{48}$Ca at 25 MeV/nucleon~and at
40 MeV/nucleon, respectively, and   Z$_{BU}$=21 for  the $^{40}$Ca case.

\begin{figure}[htb]
\includegraphics[width=0.5\textwidth]{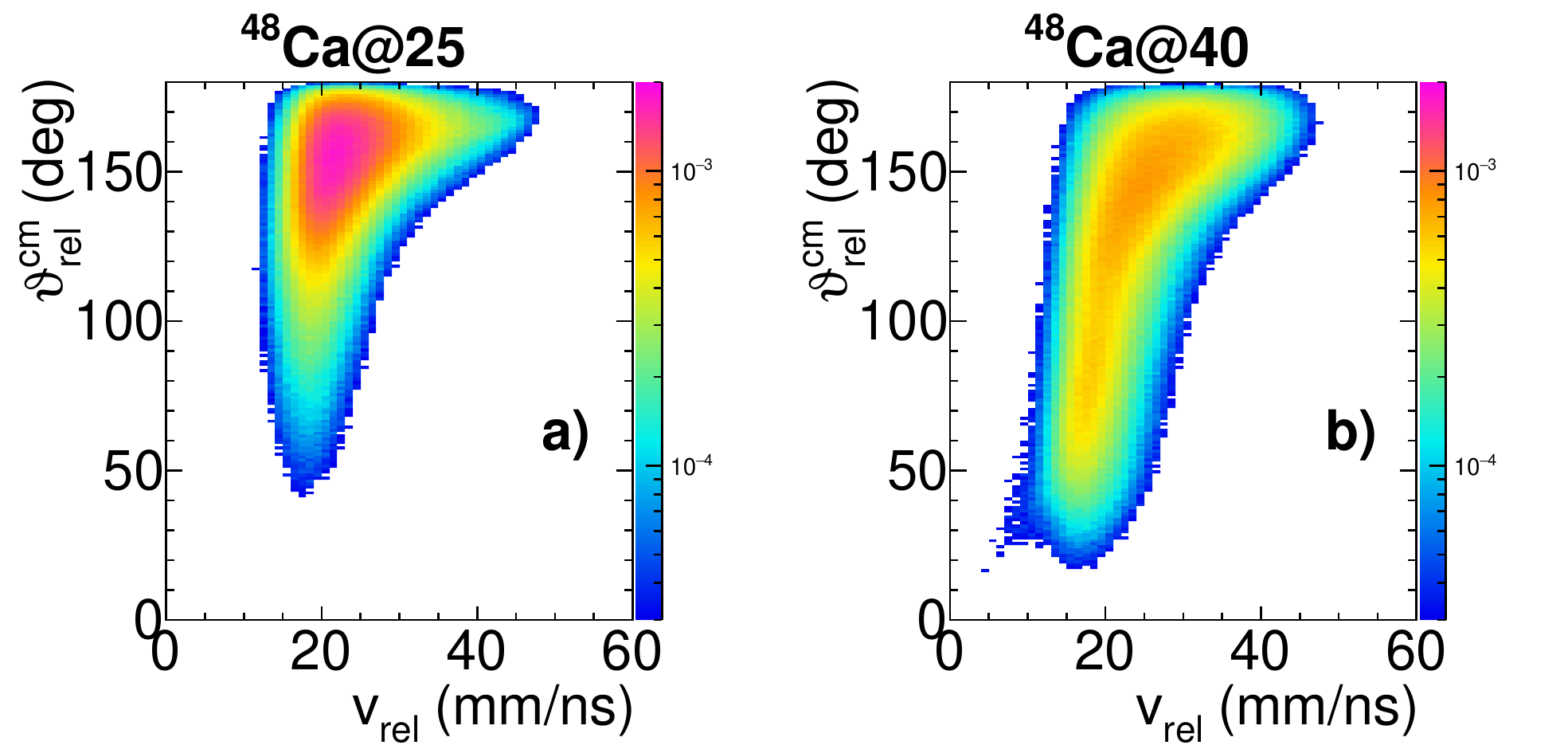}
\includegraphics[width=0.5\textwidth]{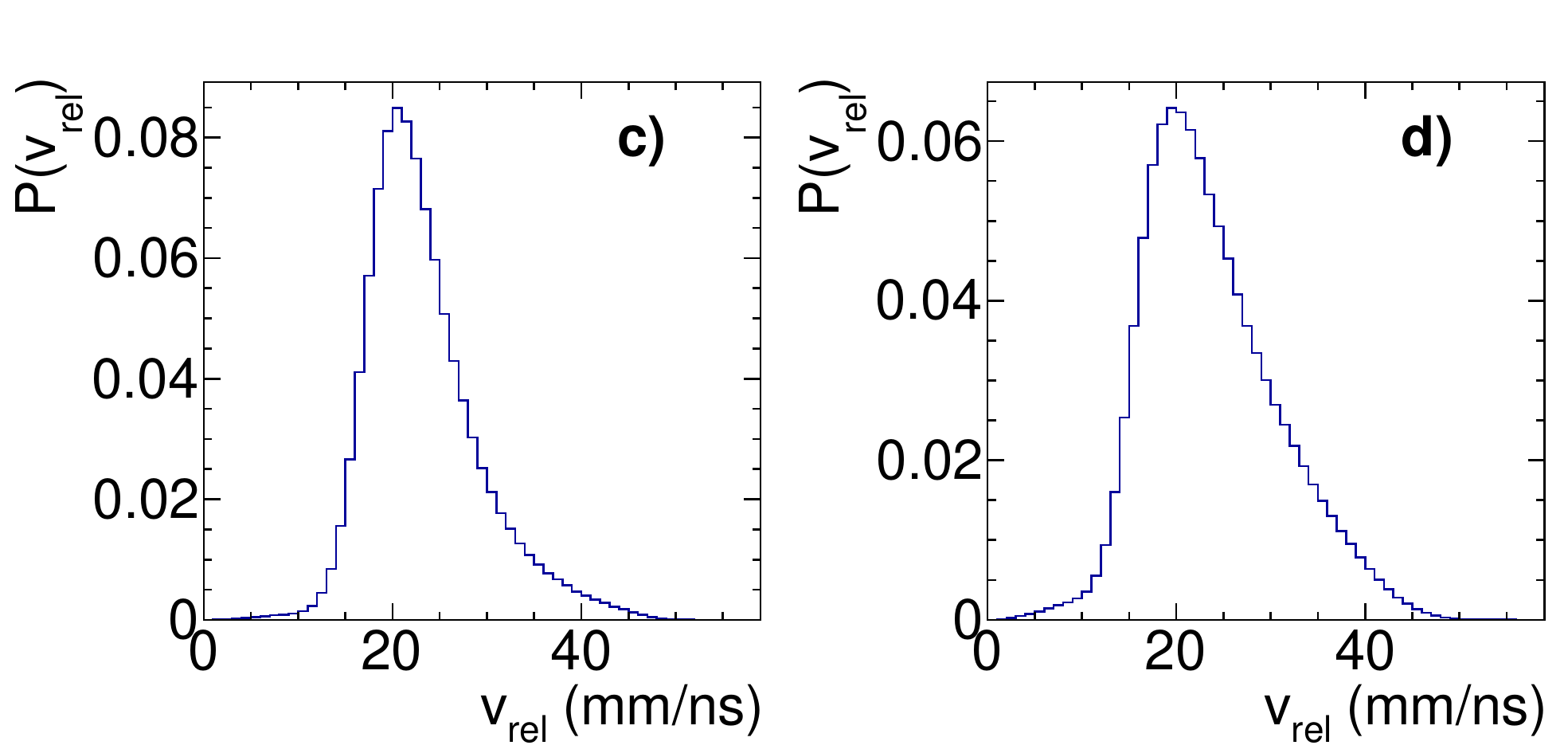}
\caption{Top: $\vartheta_{rel}^{cm}$ vs. $v_{rel}$ calculated for $HF$ and $LF$. a): $^{48}$Ca at 25 MeV/nucleon b):
  $^{48}$Ca at 40 MeV/nucleon. Bottom: Relative probability distribution of  $v_{rel}$ for the two
  reactions.} 
\label{fig5}
\end{figure}

The $\vartheta_{rel}^{cm}$ vs. $v_{rel}$ correlation (where $\vartheta_{rel}^{cm}$ is the relative angle in the c.m. frame between the two detected fragments and $v_{rel}$ is their relative velocity)
for $HF$ and $LF$ is shown in Fig. \ref{fig5} (a) and (b) for the
two reactions 
$^{48}$Ca+$^{12}$C at both beam energies;
in the panels (c) and (d) the corresponding relative velocity distributions are
drawn. 
The correlations confirm that the maximum yield is
compatible with  BU from very damped events: indeed, most 
fragments are emitted at large relative angles (essentially back-to-back) as it
is expected from a source close to $v_{c.m.}$. A tail 
extends to narrower relative angles, being more intense at the highest beam energy; this
probably corresponds to BU sources for less dissipative events, flying
at higher velocities, i.e. maybe coming from the QP breakup (for example, at 40 MeV/nucleon, the average value of the
c.m. velocity of reconstructed BUF is 40\% smaller for $\vartheta_{rel}^{cm}$ beyond 120$^{\circ}$ than  below this limit).
The different intensity of the tail at the two bombarding energies may be caused 
by 
the fact 
that at 40 MeV/nucleon the stronger kinematic
focusing favours the BU detection even for
c.m.-forward emitted
sources (i.e. closer to the beam velocity).

We thus conclude that these correlations do not clearly point out to
different event types, but they confirm that we are dealing with a
continuous and indistinct set of more or less hot primary sources,
mostly close to the 
$v_{c.m.}$ and likely heavier than the original projectile. 
In  panels (c) and (d) of Fig. \ref{fig5} the $v_{rel}$ distributions are shown.
At a first glance, we see that both distributions peak at the values 
predicted by the Viola formula \cite{Viola85}. For instance, if we consider the BU of
a representative fragment (Z$_{BU}$=20 and A$_{BU}$=44, see Table~\ref{tabaz}) with
asymmetry 0.3, the Viola formula
gives v$_{rel}=20.0$~mm/ns, very close to the maximum of the
distributions. 

A more detailed analysis on the relative velocity of $HF$ and $LF$
and its comparison with the fission systematics can be attempted by
fully exploiting the  isotopic identification of both
fission partners. For sake of clarity from now on we will consider only BU events with $\vartheta_{rel}^{cm}>120^{\circ}$, corresponding to the most dissipative reactions.

We remind that the Viola systematics describes the average kinetic
energy $E_K$ of fission fragments in Coulomb driven breakup. The systematics, initially published
for heavy 
nuclei where the fission process is more probable, was then extended
to medium mass sources. The formula gives only one value of $E_K$ for
a parent nucleus with mass A and charge Z, independently of the breakup asymmetry~\cite{Viola85}: \(E_K=0.1189Z^2/A^{1/3}+7.3\) MeV. From  $E_K$ the relative
velocity of the fragments can be obtained if both their masses are known.
In some published data for which the expected velocity for the Coulomb breakup is calculated, the two fragments are identified in charge 
~\cite{Fan2000} and
a hypothesis on the mass\footnote{In \cite{Fan2000} \(A_i=2.08Z_i+2.9\cdot10^{-3}Z_i^2\), where $Z_{i=1,2}$ is the charge of the breakup pair; moreover the relative kinetic energy is calculated as \(E_{rel}=[1.44Z_1Z_2/r_0(A_1^{1/3}+A_2^{1/3})+2.0]\)MeV with $r_0$ obtained equating $E_{rel}$ to $E_K$ of \cite{Viola85} in the hypothesis of symmetric split} is needed to obtain the relative velocity. In other
cases~\cite{Hinde1987} when measuring masses from kinematic
methods (e.g. from kinetic energy and time of flight), it is necessary  a guess on the primary
charge to derive the expected relative velocity from $E_K$ \footnote{In this case the authors explicitly introduce the split asymmetry in the calculation of the relative energy: \(E_K=[0.755Z_1Z_2/(A_1^{1/3}+A_2^{1/3})+7.3]\)MeV}. 
In this work, instead, the fragments coming from the breakup can be both isotopically
identified; therefore, at least for this subset of events, it is possible to calculate the $E_K$ given by the Viola formula using the experimentally measured Z$_{BU}$ and A$_{BU}$ on an event by event basis\footnote{i.e. \(E_K=[(0.1189)(Z_{HF}+Z_{LF})^2/(A_{HF}+A_{LF})^{1/3}+7.3]\)MeV} and the passage from $E_K$ to relative
velocity does not introduce additional hypotheses.

Nevertheless, we should note 
that since we calculate the Viola velocity ($v_{Viola}$) by means of
 quantities relative to measured (secondary) fragments, some
 effects due to the evaporation from the reseparating pair (after
 the split) might be present. However, we argue that this effect should be small because for the
nuclei under study the 
 splitting process 
 dissipates a considerable amount of energy (all the breakup reactions are
 endoenergetic in this region of masses). 

\begin{figure}[htpb]
\includegraphics[width=0.4\textwidth]{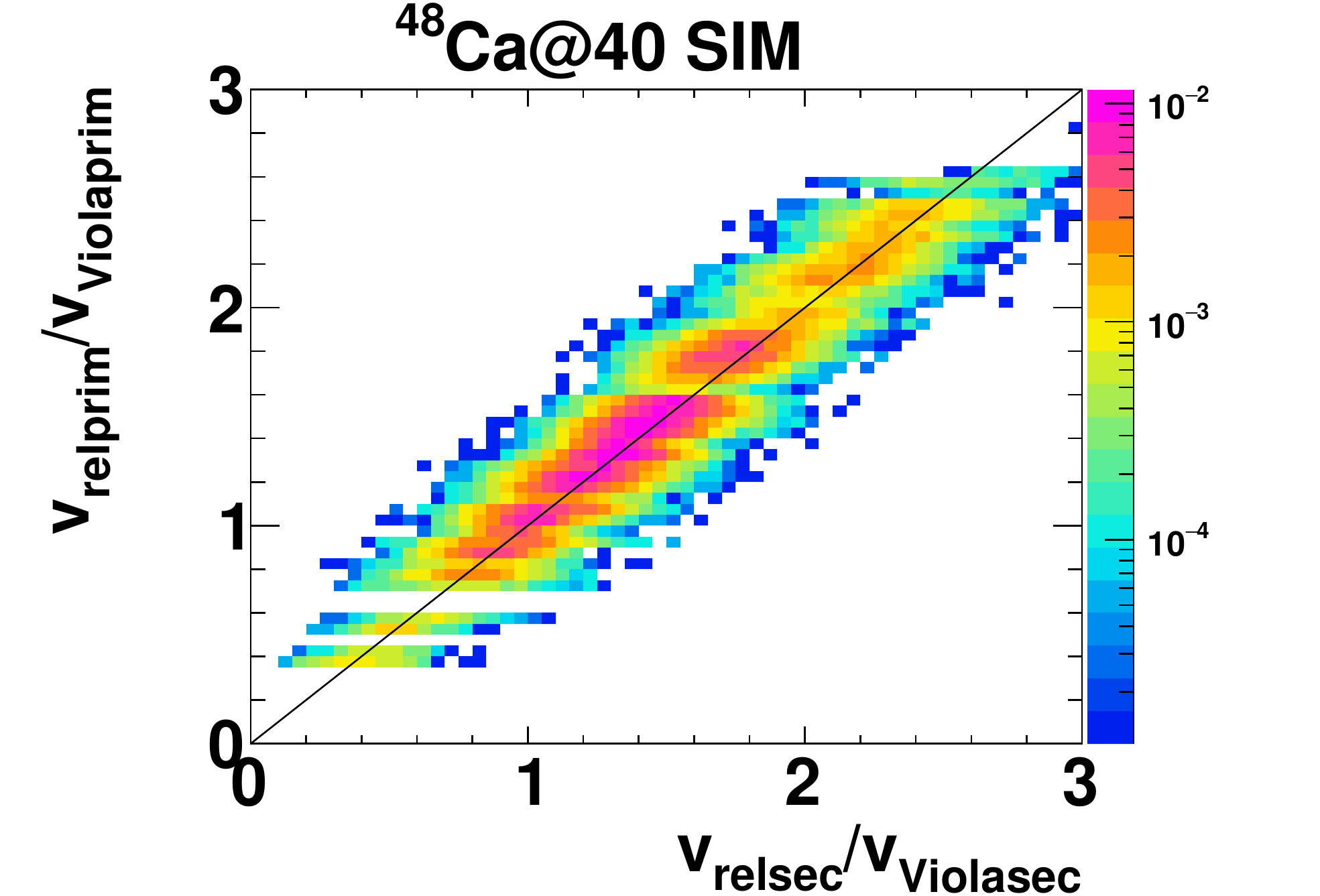}
\caption{AMD(+GEMINI++) predictions for the reaction at 40 MeV/nucleon: ratio of the relative velocity calculated at 500 fm/c to the Viola velocity (calculated with masses and charges at 500 fm/c) vs. ratio of the secondary relative velocity (after the afterburner) to the Viola velocity calculated with secondary masses and charges for BU events.} 
\label{figcorr}
 \end{figure}
\begin{figure}[htpb]
\includegraphics[width=0.4\textwidth]{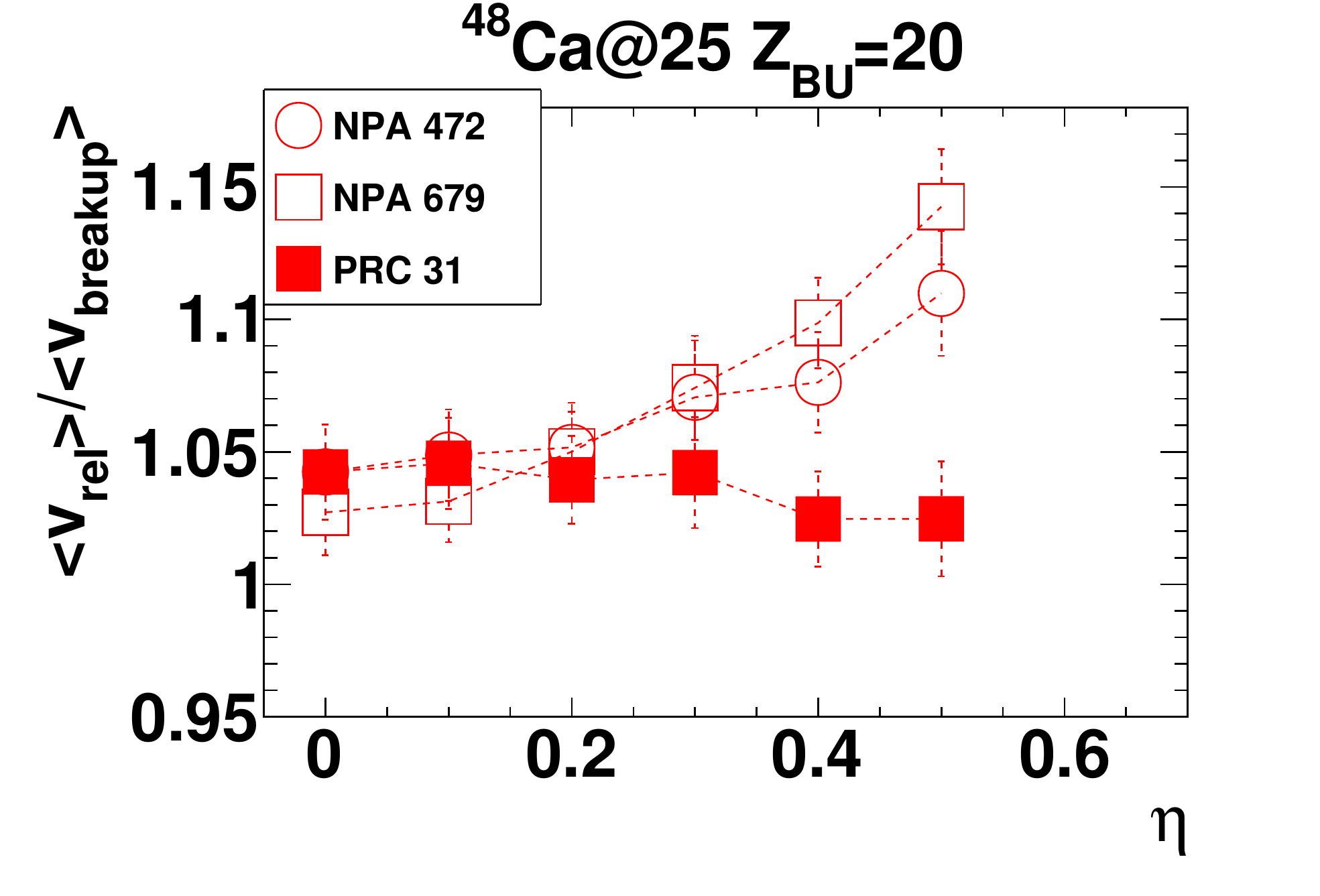}
\caption{Average relative velocity between the breakup fragments
normalized to the average velocity reported in \cite{Fan2000}[NPA679],
\cite{Hinde1987}[NPA472] and \cite{Viola85}[PRC31] as a function of the charge
asymmetry for Z$_{BU}$=20 and for the system $^{48}$Ca at 25
MeV/nucleon.} 
\label{figconfronto}
 \end{figure}

{\begin{figure*}[tp]
\begin{tabular}{ccc}
\includegraphics[width=0.33\textwidth]{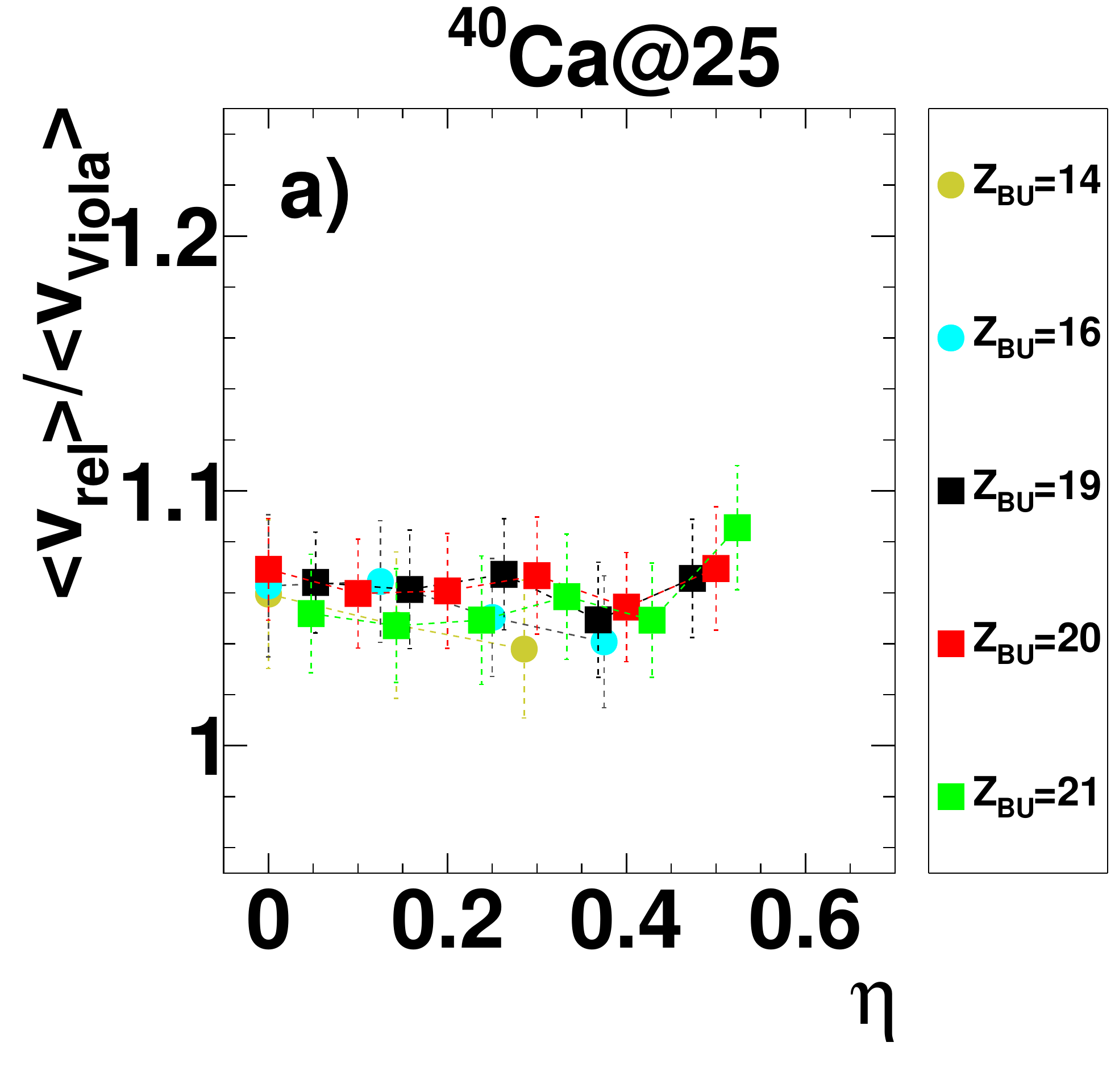}&\includegraphics[width=0.33\textwidth]{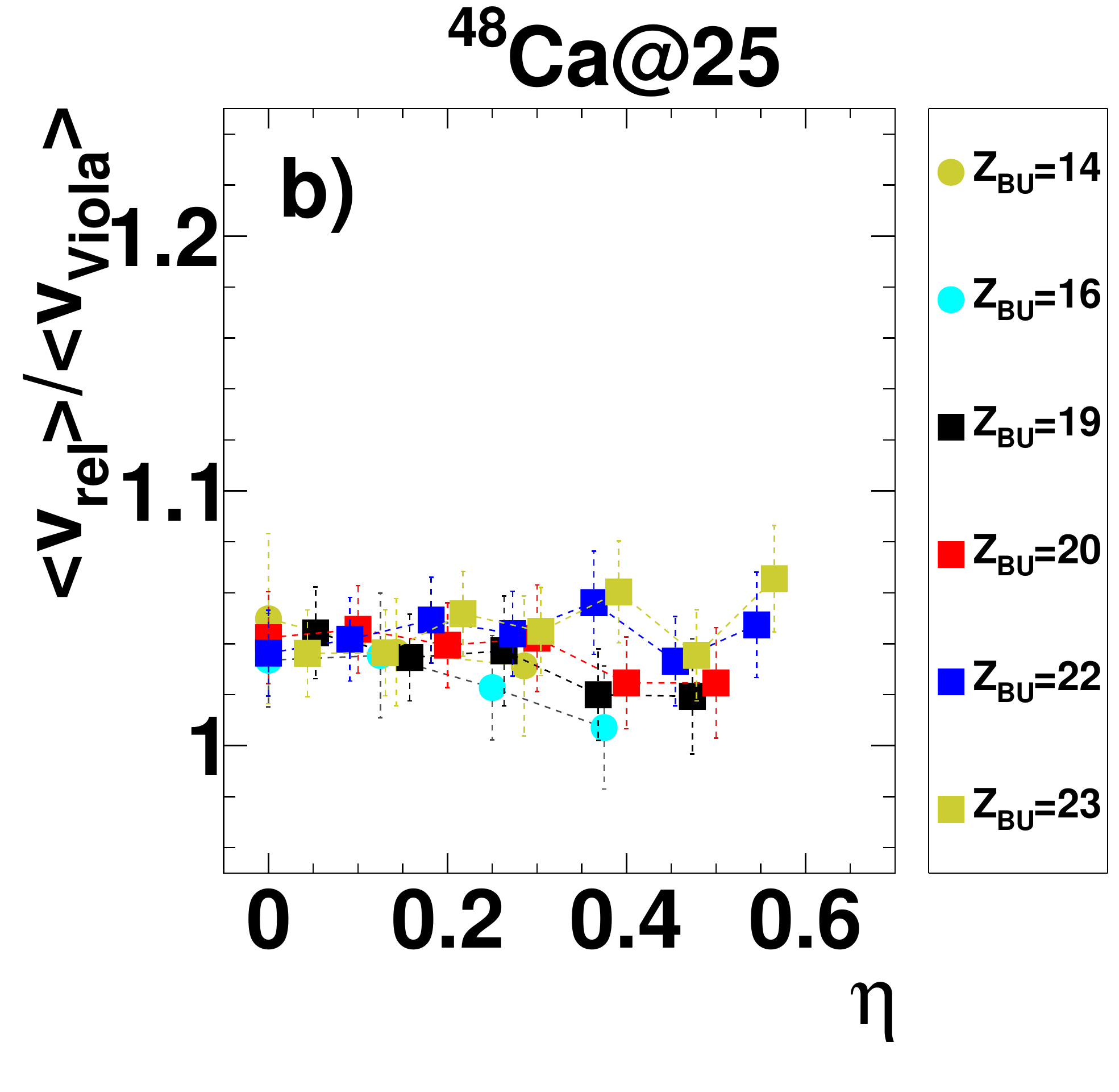}&
\includegraphics[width=0.33\textwidth]{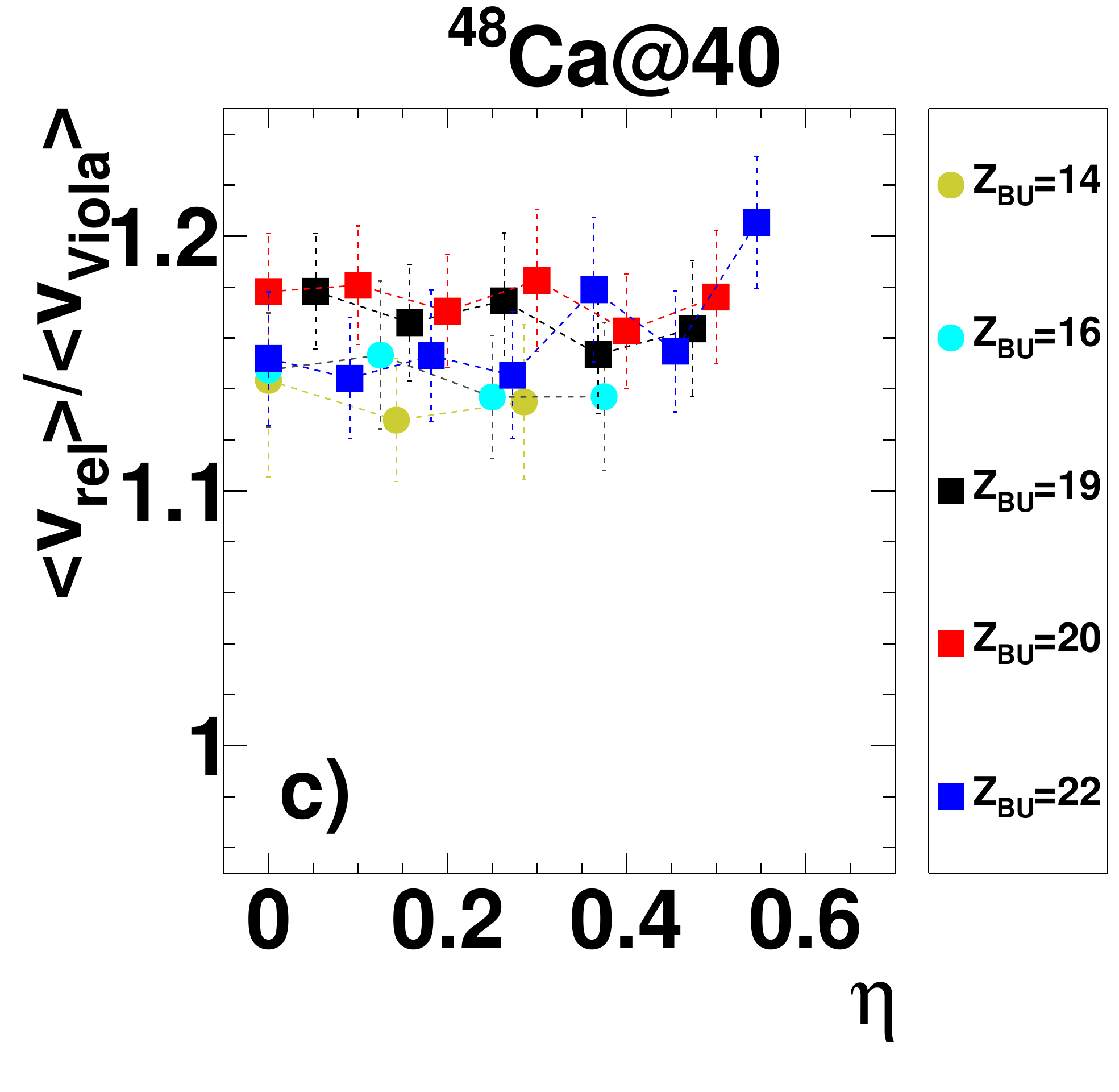}\\
\end{tabular}
\caption{Average relative velocity between $HF$ and $LF$ 
  normalized to the average $v_{Viola}$ \cite{Viola85} as a function of the charge
  asymmetry; each symbol/color corresponds to reconstructed
  fragments with charge
  Z$_{BU}$=Z$_H$+Z$_L$. Each panel shows a different system.} 
\label{fig55}
\end{figure*}

As a matter of fact, within the AMD+GEMINI++ model, we have checked this assumption: we have verified using the simulated data that when the breakup takes
 place before 500 fm/c (i.e. during the dynamical calculation) the
 ratio of the relative velocity obtained at such a time to $v_{Viola}$
 calculated with masses and charges at 500 fm/c is on
 average equal to that calculated after the evaporation using
 secondary quantities, as shown in Fig. \ref{figcorr}.

Thanks to the knowledge of mass and charge of both fission fragments, we can
check the effect of estimating the breakup velocity due to the Coulomb repulsion using the Viola formula of \cite{Viola85}  and the full experimental information (i.e. Z$_{HF}$, Z$_{LF}$, A$_{HF}$ and A$_{LF}$) or  using only partial information (only Z and the prescription of \cite{Fan2000}) and/or different formulae proposed in the literature (the formula of \cite{Hinde1987} with experimental charges and masses).
The result is shown in Fig. \ref{figconfronto} as a ratio between the experimental relative velocity and the calculated ones for Z$_{BU}$=20 exploiting the experimental data coming from the  
$^{48}$Ca low energy reaction. 
When the Viola prescription and the full experimental information are used, the obtained ratio, although greater than 1 
, is almost independent of the charge asymmetry. On the contrary, when other prescriptions and/or partial information are used, the ratio is similar to that obtained in the previous case at small charge asymmetries, while it significantly deviates from it beyond $\eta$=0.3. Therefore the use of these recipes to estimate the breakup velocity associated to the Coulomb repulsion as for example in \cite{DeFilippo2012}, where \cite{Hinde1987} is used,  may produce a distortion at large asymmetries. As a consequence in the following discussion the experimental relative velocities will be compared to those given by the Viola prescription \cite{Viola85} keeping only the events in which both breakup fragments are isotopically identified.

Figure \ref{fig55} displays the relative velocity normalized to $v_{Viola}$ for the three studied reactions as a function of the charge asymmetry and for some different Z$_{BU}$ values. For all systems the obtained ratio is almost independent of Z$_{BU}$ (the not shown cases are very similar to the displayed ones) and $\eta$; concerning the evolution with the beam energy, a clear increase moving from the systems at 25 MeV/nucleon (for which all the ratio values are between 1.04 and 1.08 for the $^{40}$Ca beam and between 1.00 and 1.06 for $^{48}$Ca) to that at 40 MeV/nucleon (where the ratios are in the range 1.13-1.20) is observed.
%(ratio around 1.06$\pm$2\% for the $^{40}$Ca beam and around 1.04$\pm$2\% for $^{48}$Ca) to that at 40 MeV/nucleon (ratio around 1.16$\pm$3\%) is observed.

This finding suggests
the occurrence of some additional contributions (beyond Coulomb
repulsion) acting during  the breakup process, whose weight increases with the beam energy; possible explanations of this effect might be
the larger angular momentum of the system attained during the evolution of the dynamical phase and/or a more stretched phase space configuration of the breaking source.

Concerning the angular momentum, a rough calculation
indicates that the ratio $\mathcal{R}$ of the relative velocity between $HF$ and $LF$  and the velocity due to Coulomb repulsion
(calculated from
\(E_{Coul}=1.44\frac{Z_HZ_L}{(1.2\,A_H^{1/3}+1.2\,A_L^{1/3})}\)
MeV) should depend on the angular momentum of the system according to the
formula \(\mathcal{R}=\sqrt{1+\frac{E_{rot}}{E_{Coul}}}\), where
\(E_{rot}=\frac{L^2}{2I}\) is the rotational energy. In this formula
$L$  is the angular momentum and $I$ is the moment of inertia calculated
for two rigid spheres (corresponding to the BU fragments) in contact with respect to a perpendicular axis passing through
their c.m.\footnote{\(I=\frac{2}{5}[A_H\cdot(1.2\,A_H^{1/3})^2+A_L\cdot(1.2\,A_L^{1/3})^2]+A_{red}(1.2\,A_H^{1/3}+1.2\,A_L^{1/3})^2\),
where \(A_{red}=\frac{A_LA_H}{(A_L+A_H)}\)}. 
Figure \ref{figL} shows the
expected ratio $\mathcal{R}$ as a function of $\eta$ for different values of $L$ for Z$_{BU}$=20. For $L=0$ the ratio is exactly 1 by construction
and it grows of about 4-5\% for an increase of  20$\hbar$. Therefore, if we suppose that the average angular momentum of the
fragmenting system increases with the beam energy, we can expect an increase of the $v_{rel}/v_{Viola}$ ratio.

Although the AMD simulation does not guarantee angular momentum conservation if not purposely forced \cite{Piantelli2020}, we have verified that within this model the hypothesis that the average angular momentum of the breaking up fragment increases with the beam energy is not supported. In fact for all the simulated reactions the spin distributions of the primary fragments just before the breakup taking place during the dynamical phase are very similar and with average values below 20$\hbar$.

Nevertheless, the simulation AMD+GEMINI++ still predicts an increase of the $v_{rel}/v_{Viola}$ ratio with the beam energy (variation of the order of 20\% from 25 MeV/nucleon to 40 MeV/nucleon) as in the experimental case, as shown in
Fig.\ref{vrapsim} for some Z$_{BU}$, although in this case a systematic decrease of the ratio for all Z$_{BU}$ when $\eta$ increases appears, at variance with the experimental data; moreover, the ratio at 25 MeV/nucleon is always below 1.

Looking at the whole $v_{rel}/v_{Viola}$ distribution it is possible to put into evidence that both in the experimental and in the simulated case the difference in the average value between 25 MeV/nucleon and 40 MeV/nucleon is mainly due to a tail extending towards high values in the case of the high energy beam; this behaviour may reflect the fact that at 40 MeV/nucleon the breaking up fragment is more stretched due to dynamical effects, i.e. it has a more elongated shape (in velocity) due to a stronger memory of the entrance channel. As a  support of this hypothesis we verified  that in the model the velocity distribution along the beam axis of the single nucleons forming the fragment just before the breakup is broader at 40 MeV/nucleon than at 25 MeV/nucleon.

\begin{figure}[htpb]
\includegraphics[width=0.33\textwidth]{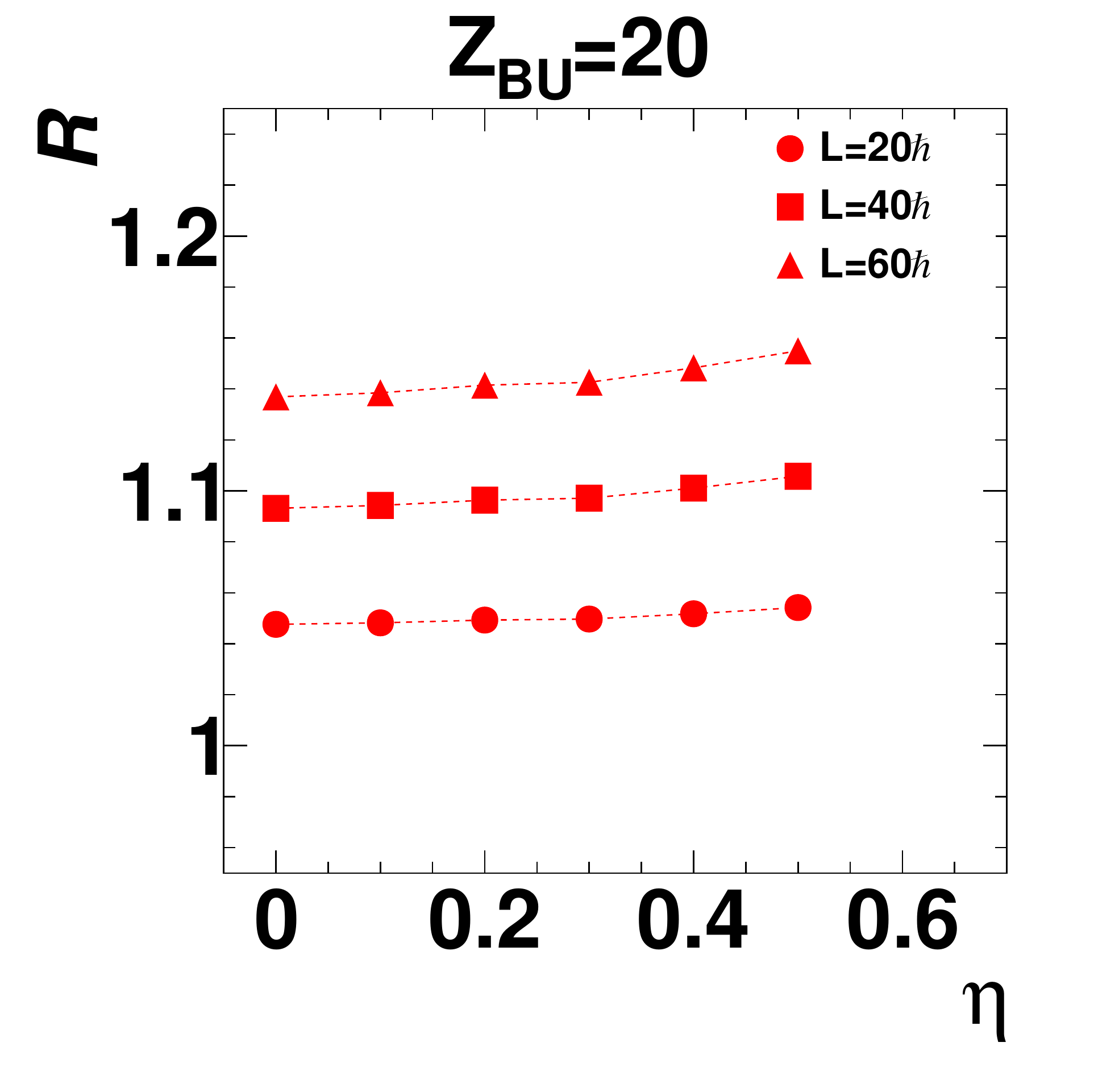}
\caption{\(\mathcal{R}=\sqrt{1+\frac{E_{rot}}{E_{Coul}}}\) calculated for different values of angular momentum of the breaking up system as a function of $\eta$ for $Z_{BU}$=20.}
\label{figL}
\end{figure}

\begin{figure}[htpb]
\begin{tabular}{cc}
\includegraphics[width=0.25\textwidth]{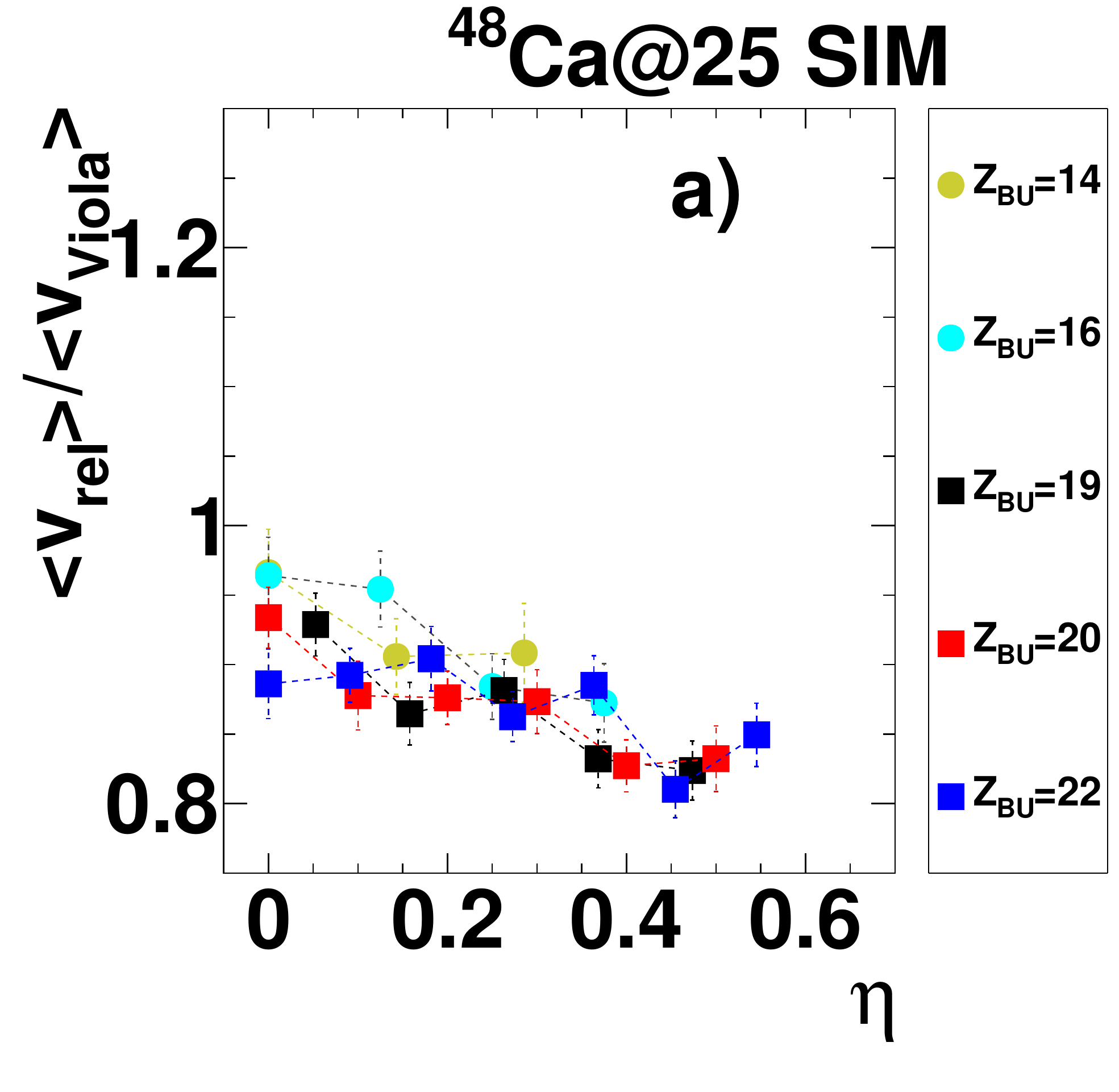}&\includegraphics[width=0.25\textwidth]{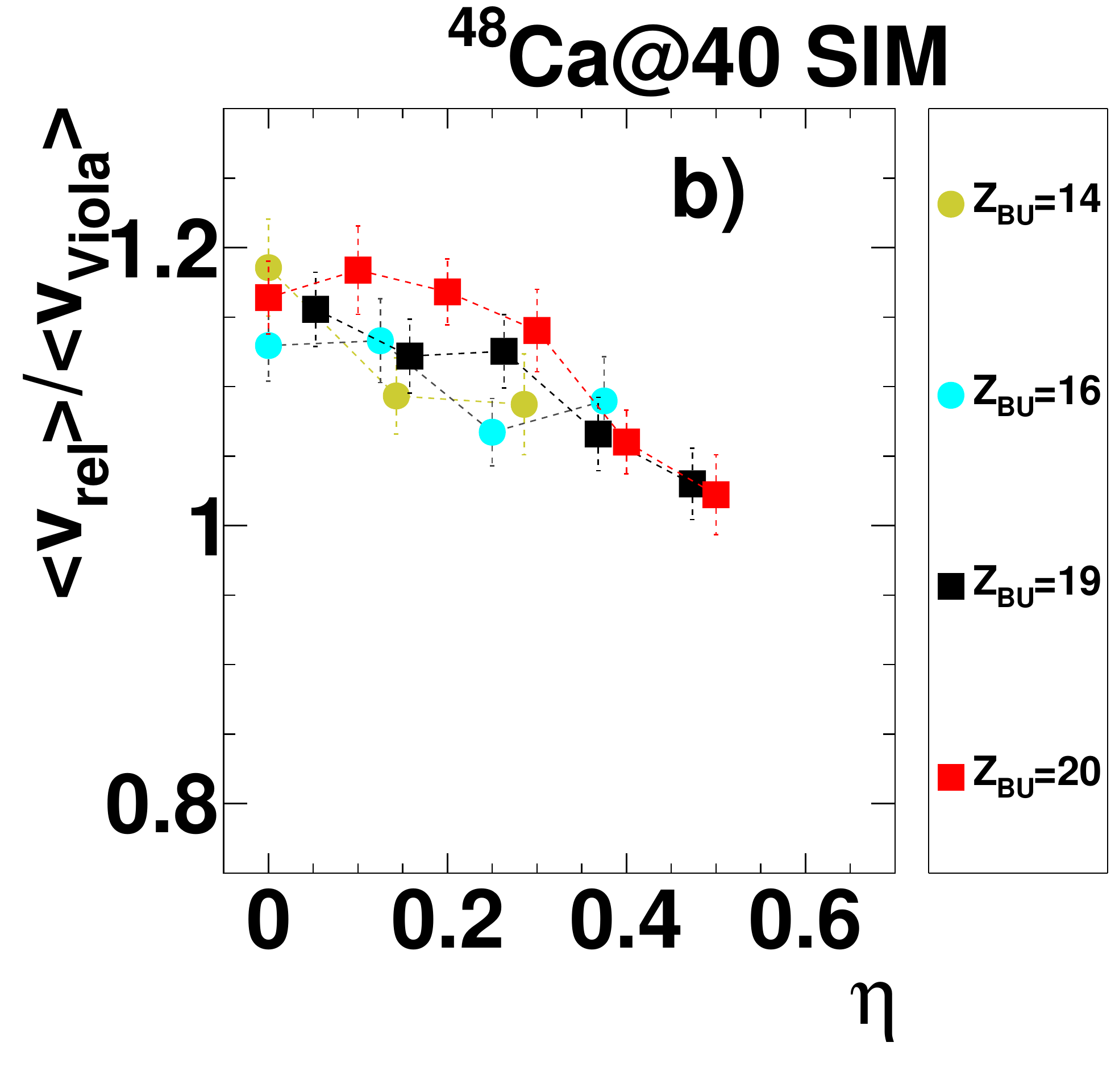}\\
\end{tabular}
\caption{Simulated data (AMD+GEMINI++): average relative velocity between $HF$ and $LF$ normalized to the average $v_{Viola}$ as a function
  of the charge asymmetry; each symbol/color corresponds to a
  different 
Z$_{BU}$ (same legend as in Fig. \ref{fig55}). Each panel shows a different system.} 
\label{vrapsim}
\end{figure}

Therefore these evidences suggest the interpretation of the trend observed in Fig. \ref{fig55} (b) and (c) in terms of an increasing weight of dynamical effects.
Anyhow, despite such a contribution, relative velocity distributions located close to the values coming from the fission systematics are a signature of a basically Coulomb-driven breakup. Therefore in the following we will use such an observable as a probe of the fact that also a fragment reconstructed via particle correlation and detected in coincidence with a heavier one in many cases can be interpreted as the $LF$ of a breakup process.

\begin{figure}
\includegraphics[width=0.5\textwidth]{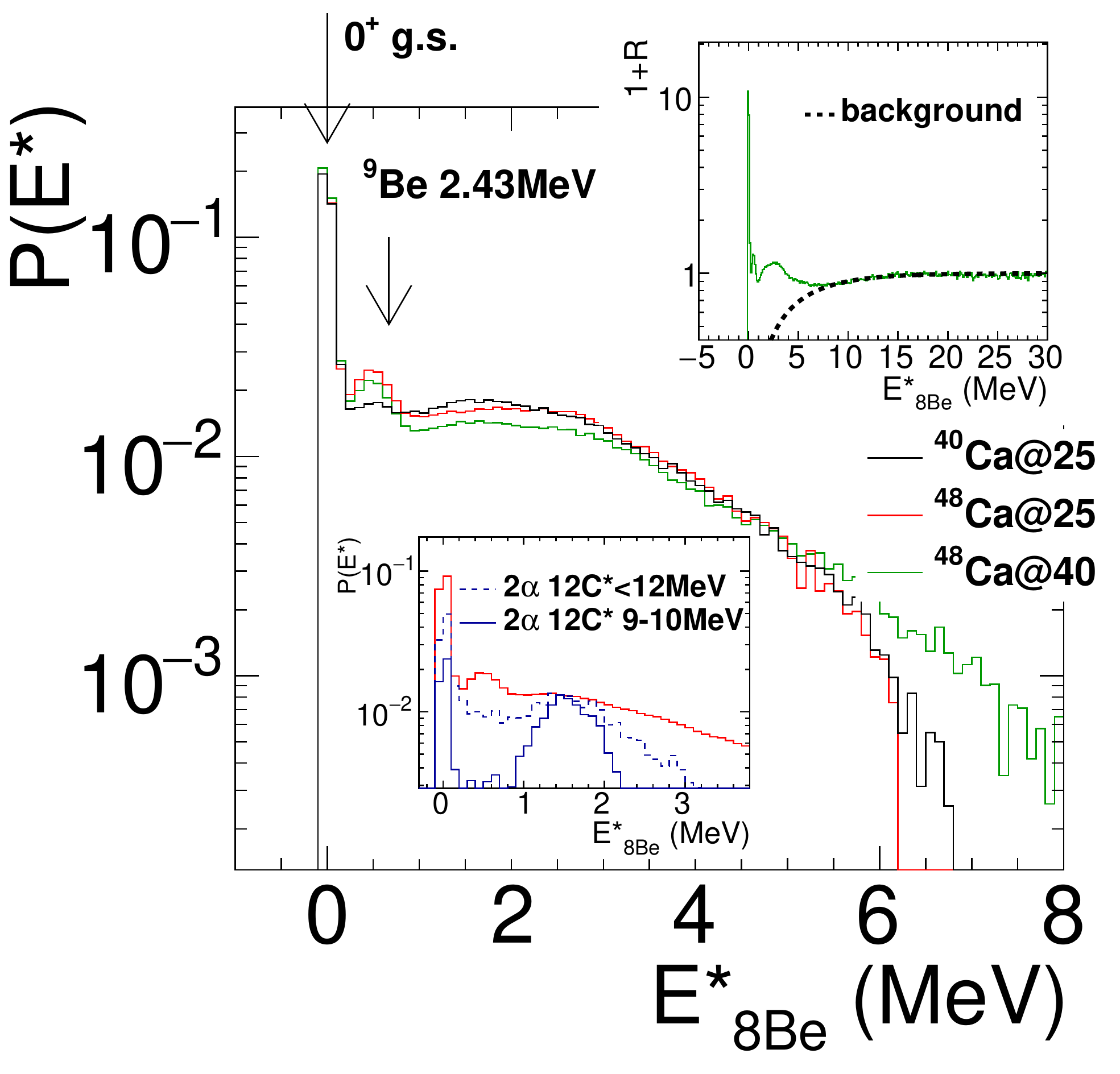}

\caption{Relative probability ditribution of the excitation energy of $^8$Be$^{*}$ reconstructed from 2
  $\alpha$ correlations, after background subtraction. In the top inset  1+R(E$^{*}$) vs. E$^{*}$
  for $^{48}$Ca  at 40 MeV/nucleon is shown as green line; the black dashed
  line is the Coulomb background. Bottom inset: relative probability distribution of the $^8$Be$^{*}$ excitation
  energy spectrum without background subtraction for the system $^{48}$Ca at 25
  MeV/nucleon (red continuous line); dashed lines: feeding from $^{12}$C$^{*}$  with $E^{*}<12$MeV (details in the text). Continuous line: the same but with $E^{*}$ of the $^{12}$C$^{*}$ between 9 MeV and 10 MeV.
The normalization is done at
  1.4MeV. }  
\label{8Be} 
\end{figure}
\subsubsection{Breakup channel with the light fragment  recontructed via particle fragment correlations}

In the BU process the excitation energy of the splitting fragment can be high enough to allow for further particle decay. In particular if the $LF$ is produced in an excited state decaying via particle emission and if its decay products are detected in coincidence, the $LF$ can be reconstructed by means of the particle-particle correlation technique.

The 
correlation technique has been  widely used in the literature to
investigate the emission source size and/or the emission time
\cite{Verde1,Verde2}. In this paper we mainly restrict the analysis to the resonances of C isotopes decaying in channels including the emission of $\alpha$ particles since they are collected with
high statistics.   The technique
basically consists in 
selecting events where two particles (labelled as 1 and 2), candidates to be 
the decay products of a parent resonance, are detected. We build the observable 
\(1+R(E_{rel})=C\frac{\sum{Y_{12}(\vec{p_1},\vec{p_2})}}{\sum{Y_{12}^{unc}(\vec{p_1},\vec{p_2})}}\),
where \(Y_{12}(\vec{p_1},\vec{p_2})\) are the
correlated yields of the pair (i.e. the pairs are detected in the same
event) and $\vec{p_i}$ with $i=1,2$ is the particle momentum,
while \(Y_{12}^{unc}(\vec{p_1},\vec{p_2})\) are the uncorrelated ones,
estimated by means of the event mixing technique \cite{Drijard}. The summations run over pairs of momenta $\vec{p_1}$ and $\vec{p_2}$ corresponding to the same value of relative energy $E_{rel}$, where \(E_{rel}=(\vec{p_1}-\vec{p_2})^2/2\mu\), with $\mu$ the reduced mass of the pair. $C$ is a
normalization coefficient estimated at large $E_{rel}$
with the condition $R(E_{rel})\rightarrow 0$. The building of the
uncorrelated yields is 
a critical procedure  which should take
into account the efficiency of the setup for the different kinds
of events, the event class selection (which should be the same both for the
correlated and the uncorrelated case), the conservation laws (which
should be checked not to be violated for uncorrelated events, otherwise distortions can be produced)\cite{Chajecki,Verde3,Tan}.
We also note that in the present case the previous issues for
  proper background 
evaluation are enhanced by the limited angular coverage of the
apparatus
which in addition differently affects the efficiency for the
reactions  at 25 and 40 MeV/nucleon. Therefore the following analysis
is focused on showing that the emission pattern of the reconstructed fragments is compatible with that of the $LF$ in BU events,
 without attempting quantitative analysis on the population of
the different levels, also because many of them are not resolved. In
fact the setup is mainly limited by the angular resolution of the telescopes
($\Delta\vartheta$ around 1.26$^\circ$ in the adopted configuration,
where $\vartheta$ is the polar angle) and to a lesser extent by the
energy resolution (around 0.1 MeV). More
  quantitiative studies on particle particle correlations may be the
  subject of future investigations with the complete setup
  INDRA-FAZIA, where the topic of the energy and momentum conservation
  effects on the event mixing spectrum will be quantitatively faced
  thanks to the very high angular coverage of the device.

\paragraph{{\bf$^{12}$C$^*$}}

Concerning $^{12}$C, reconstructed through its decay in 3 $\alpha$ (particularly interesting as a testbench of clustering in nuclei~\cite{borderie_sym_2021,borderie_plb_2016,AquilaPRL}), we limit the analysis to the cases in which the decay is of sequential type (predominant for the Hoyle state \cite{Morelli_2016,AquilaPRL}) and proceeds through the formation of $^{8}$Be. 
Such lighter cluster is reconstructed starting from 2 $\alpha$ correlations; the obtained excitation energy spectra for the three reactions (after the Coulomb background subtraction), normalized to their integral to better compare the shapes, are shown in the main frame of Fig. \ref{8Be}. These spectra refer to all detected events with at least 2 $\alpha$, without specific selections. In the top inset of Fig. \ref{8Be} the  \((1+R(E^*))\) vs. $E^*$ distribution\footnote{\(E^{*}=E_{rel}-Q_{value}\)}, which is the starting point used to extract the excitation energy spectrum, is drawn for the system at
40 MeV/nucleon; in the plot also the Coulomb background, fitted
in the smooth region 
$E^{*}\ge20$ MeV according to the formula 
\(y=1-e^{-E^{*}/E_c}\) \cite{Nayak} ($E_c$ is the fit parameter), is shown.

The excitation energy spectra depicted in the main frame of Fig. \ref{8Be}  clearly show the narrow peak close to zero
corresponding to the ground state of $^{8}$Be (91~keV); 
a smaller peak, more abundant
for the $^{48}$Ca systems, is located around 0.5 MeV. As discussed in the literature~\cite{Pochodzalla} it
mainly corresponds to the incomplete reconstruction of 
the excited $^{9}$Be (energy level: 2.43 MeV  
\footnote{The obtained value around 0.5 MeV is slightly below the
  expected one of 0.67 MeV; the expected  value is calculated from
  \(E^{*}=E^{*}_{^9Be}
  -Q_{^9Be\rightarrow^8Be+n}-Q_{^8Be\rightarrow\alpha+\alpha}\). This shift may be due to a deformation of the excitation energy spectrum caused by a defective production of the uncorrelated yields.})  
for which the free neutron is not detected; 
it is reasonable that its contribution is larger  
for the neutron rich \cb~reactions; its population also
decreases passing from 25 to 40 MeV/nucleon\footnote{The decay of
$^{9}$Be from the level at 1.68 MeV (with 100\% probability of neutron
decay) falls in the region of the $^{8}$Be ground state and it cannot be resolved, thus acting as a further background for this state.}. The presence of the second excited state 2$^+$ at 3.03~MeV of $^{8}$Be is masked by the background due to incompletely reconstructed $^{12}$C$^{*}$ nuclei decaying
into 3 $\alpha$, as shown in the bottom inset of Fig. \ref{8Be} for the system  $^{48}$Ca at 25 MeV/nucleon. In this plot the excitation energy of $^8$Be without background subtraction (continuous red line) is presented for events with at least 2 $\alpha$, together with the spectra obtained rejecting one $\alpha$ in events with 3 detected $\alpha$ and excitation energy of the $^{12}$C$^{*}$ (reconstructed as explained in the following and shown in Fig.~\ref{12C} panel (a)) in the region $E^{*}<12$~MeV (dashed line) or for $9$ MeV$<E^{*}<10$ MeV (continuous blue line). By the way, as already shown in \cite{Manfredi2012}, this plot suggests that also the resonances of $^{12}$C around 9-10 MeV tends to decay via the $^{8}$Be, as the Hoyle state; in fact, rejecting one $\alpha$, the ground state of the $^{8}$Be is clearly populated.

From now on we gate on the ground state of $^{8}$Be and we ask for the coincidence with an heavy fragment (Z$\geq$10). 
A third coincident $\alpha$ particle is then required and the excitation energy spectrum of $^{12}$C$^{*}$ shown in Fig. \ref{12C} (a) is produced\footnote{\(E^{*}_{12C}=\sum_{i=1}^{3}{E_i-Q_{^{12}C\rightarrow3\alpha}}\), where $E_i$ is the kinetic energy of the $i^{th}$ $\alpha$ calculated with respect to the c.m. of the 3 $\alpha$} for all the reactions (the histograms  are normalized to their integral). In this case the  exclusive requirements imposed to 
produce such a spectrum (a 4-body channel, 3 $\alpha$ and a heavy fragment) strongly reduce 
the uncorrelated background; therefore it is not strictly 
necessary to divide for the
uncorrelated yields and to subtract the background at least to give
 qualitative interpretation of these events.

 \begin{figure*}[htpb]
\begin{tabular}{ccc}
\includegraphics[width=0.33\textwidth]{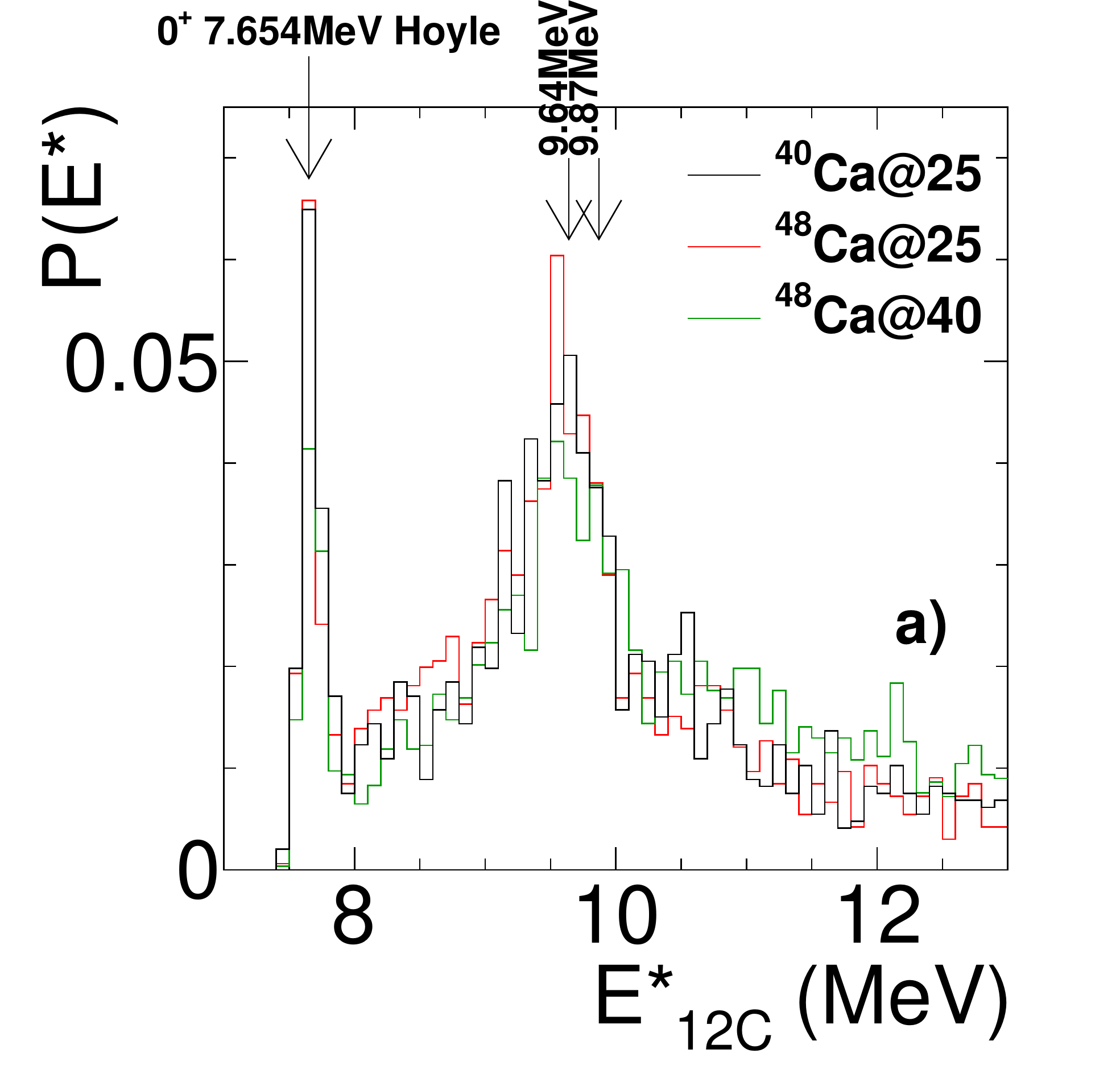}&\includegraphics[width=0.33\textwidth]{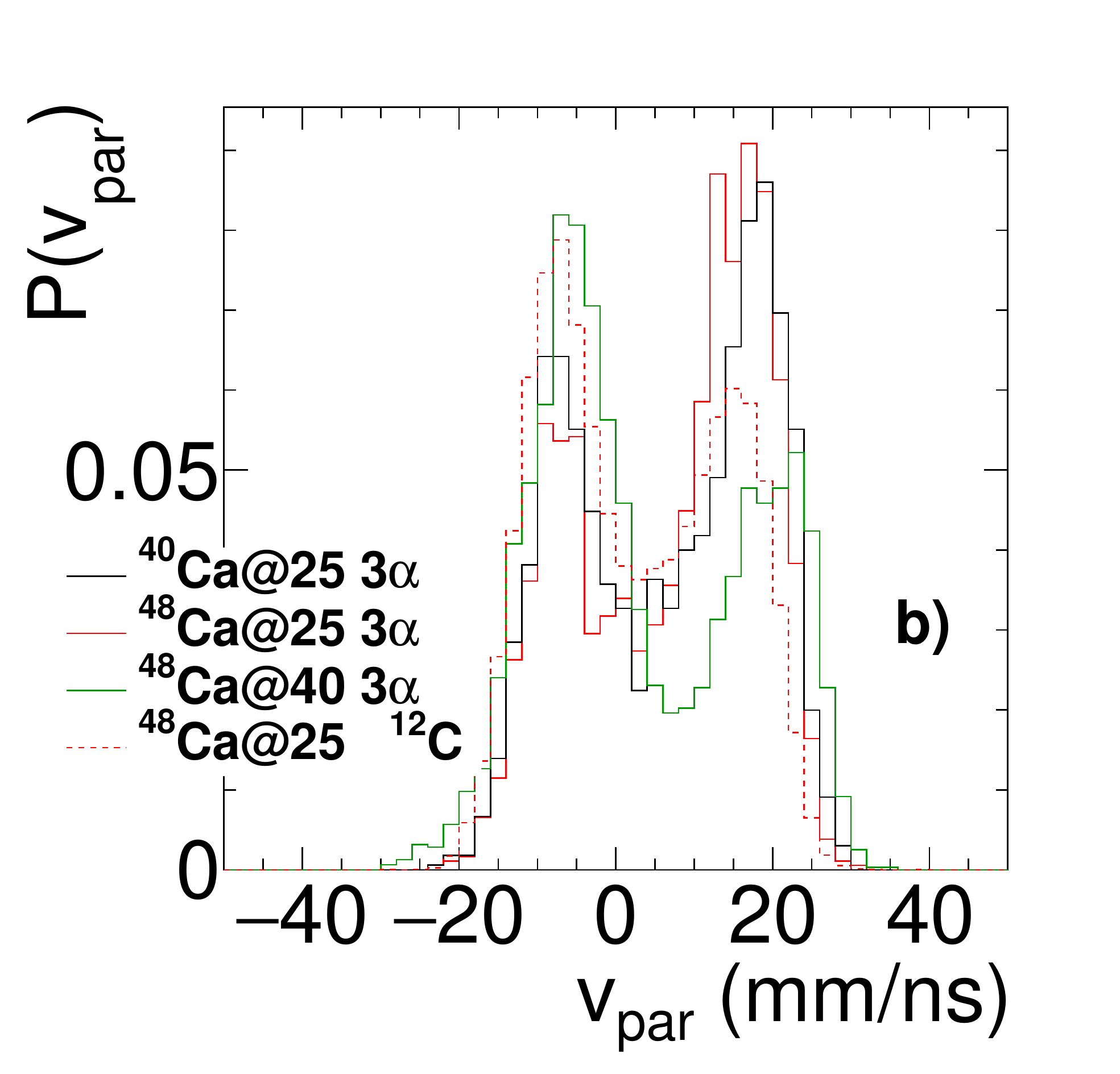}&\includegraphics[width=0.33\textwidth]{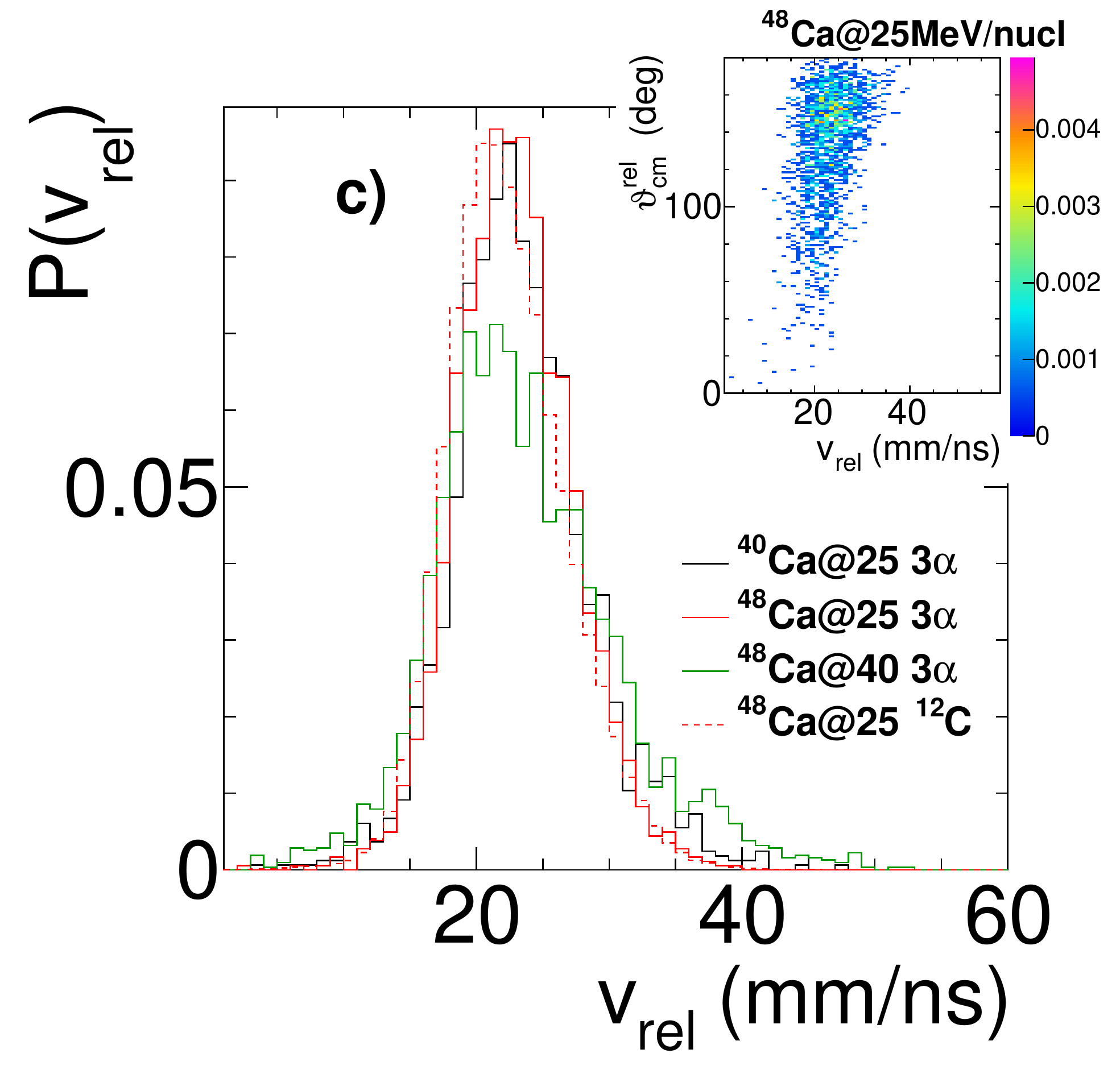}\\
\end{tabular} 
\caption{a): Relative probability distribution of the excitation energy of $^{12}$C$^{*}$ reconstructed from
  3-$\alpha$ correlations, two of them ascribed to 
  $^{8}$Be ground state, in events with  M$_{\alpha}=3$  and with a 
  fragment with Z$\ge$10 in coincidence. No background
  subtraction nor uncorrelated yield normalization have been
  applied. Some relevant excited levels are marked with arrows. b): Continuous histograms: relative probability distribution of the c.m. parallel
  velocity of the reconstructed $^{12}$C$^{*}$ calculated with respect
  to the velocity of the biggest fragment of the event, corrected for the correlated recoil, for
  $E^{*}_{12C}<$15 MeV. Dashed histogram: the same but for the detected (cold)  $^{12}$C. c): Continuous histograms: relative probability distribution of the relative velocity
  between the  reconstructed $^{12}$C$^{*}$  and the big
  fragment, for $E^{*}_{12C}<$15 MeV. Dashed histogram: the same but
  for the detected (cold) $^{12}$C. In the inset, plot of the
  relative c.m angle vs. relative velocity between the
  $^{12}$C$^{*}$ and the big fragment for the system   $^{48}$Ca at 25 MeV/nucleon{}.}  
\label{12C}
\end{figure*}
The spectra are very similar for the 3 reactions.
The first narrow peak at 7.65 MeV is the Hoyle state; a broader peak 
clearly emerges below 10~MeV corresponding to known (unresolved) states
around 9.6-10~MeV. 
In more detail the yield ratio of the highest energy peaks (between 8.85 MeV and 10.25 MeV) to the Hoyle peak (between 7.45 MeV and 7.95 MeV) shows a trend to increase with the beam energy (from 3.5$\pm$0.3 for $^{48}$Ca at 25 MeV/nucleon to 3.9$\pm$0.4 for the same reaction at 40 MeV/nucleon), thus suggesting that higher excitations may be favoured by larger beam energies. The neutron poor system gives 3.1$\pm$0.3, perhaps meaning that final states closer to the g.s. of  $^{12}$C can be favoured in the N=Z system.

The c.m. parallel velocity spectra of the reconstructed $^{12}$C
resonances, calculated with 
respect to the direction of the heaviest fragment of the event,
are shown in Fig. \ref{12C} (b) for $E^{*}_{12C}<$15 MeV.
For all the systems the distribution is almost symmetric with respect to the heavy fragment velocity, which is close to the c.m.;
the two reactions at 25 MeV/nucleon~produce very similar
distributions, while for the high energy reaction the forward peak is slightly shifted towards higher velocities.
Such an emission pattern is compatible with the presence of an almost
fully damped source and, anyhow, it is similar to the parallel velocity distribution obtained when the $^{12}$C is detected as such 
(dashed line in Fig. \ref{12C} (b)).
The residual shape difference might be due to the efficiency variation for the detection of a $^{12}$C as such and of 3 $\alpha$ in coincidence.

The coincidence between a Carbon (detected or reconstructed) and an
heavy fragment (playing the role of $HF$) belongs to the BU channel as
it was defined in this work (Z$_{LF}\geq$5); therefore it is worth
looking at the relative velocity between the reconstructed
$^{12}$C$^*$ and the coincident heavy fragment (Fig. \ref{12C} panel
(c)). The $v_{rel}$ distribution obtained for all the reactions is
compatible with a mainly Coulomb-driven breakup process with peak
values around 21~mm/ns, in close similarity with the distribution of
Fig. \ref{fig5} panel (c). Also the $\vartheta_{rel}^{cm}$ vs. $v_{rel}$
correlation, shown in the inset, remarkably looks like that of
Fig. \ref{fig5} panel (a). For $^{48}$Ca at 25 MeV/nucleon also the
$v_{rel}$ distribution obtained when the $^{12}$C is detected as such
is shown as red dashed line in the main frame and it results to be
very similar to the reconstructed one (red continuous line).

\begin{figure*}[tpb]
\begin{tabular}{ccc}
\includegraphics[width=0.33\textwidth]{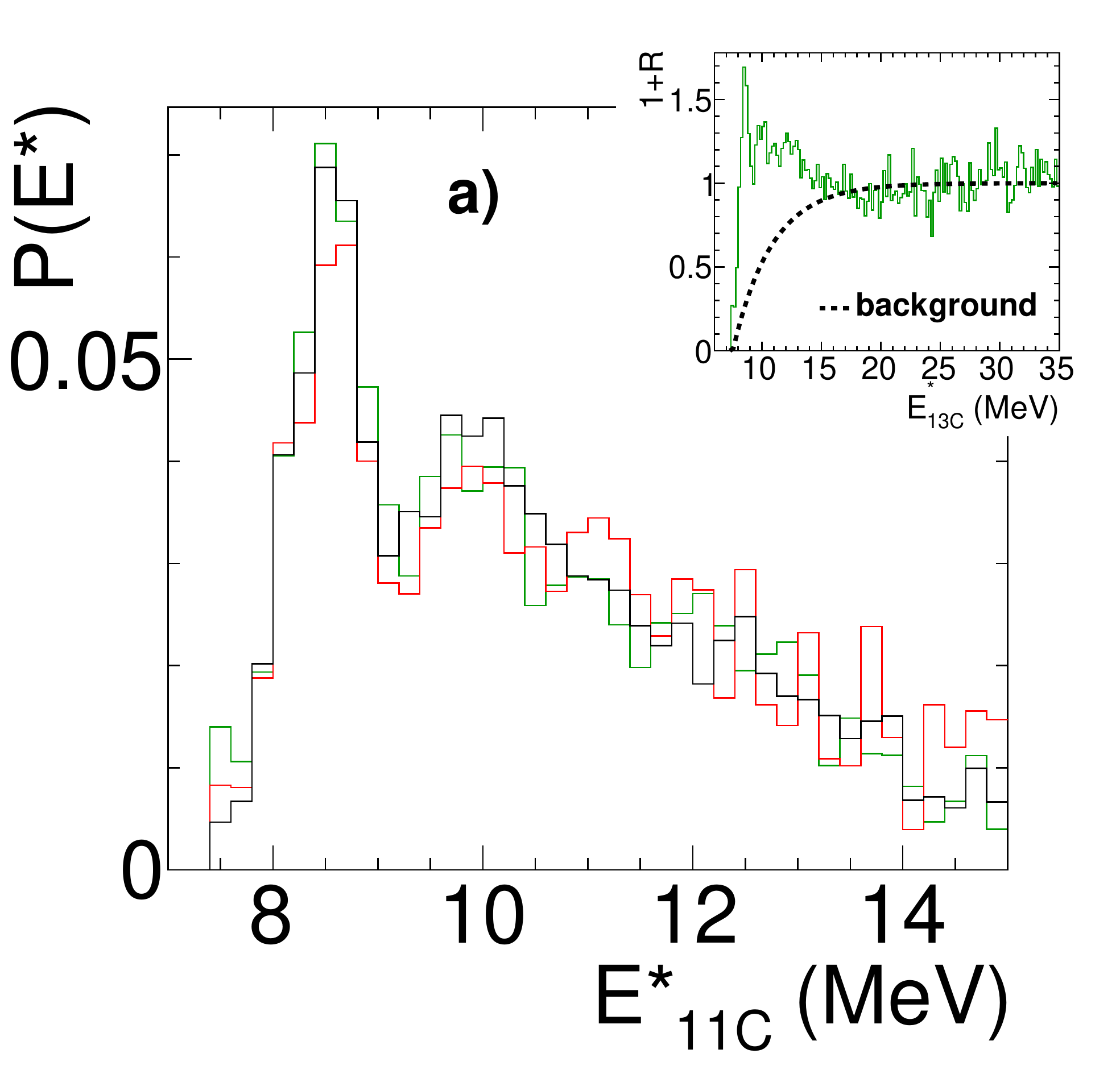}&\includegraphics[width=0.33\textwidth]{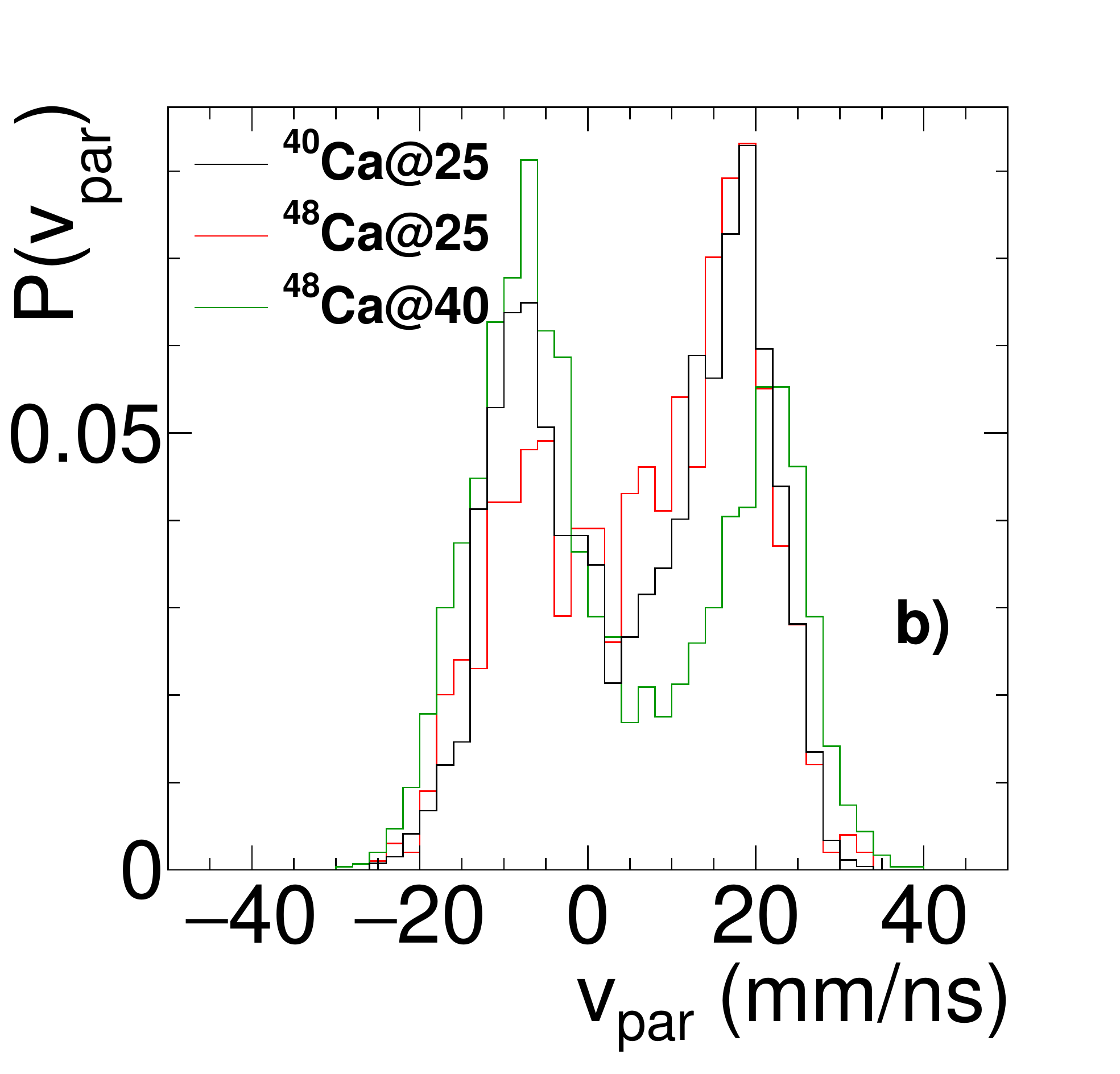}&\includegraphics[width=0.33\textwidth]{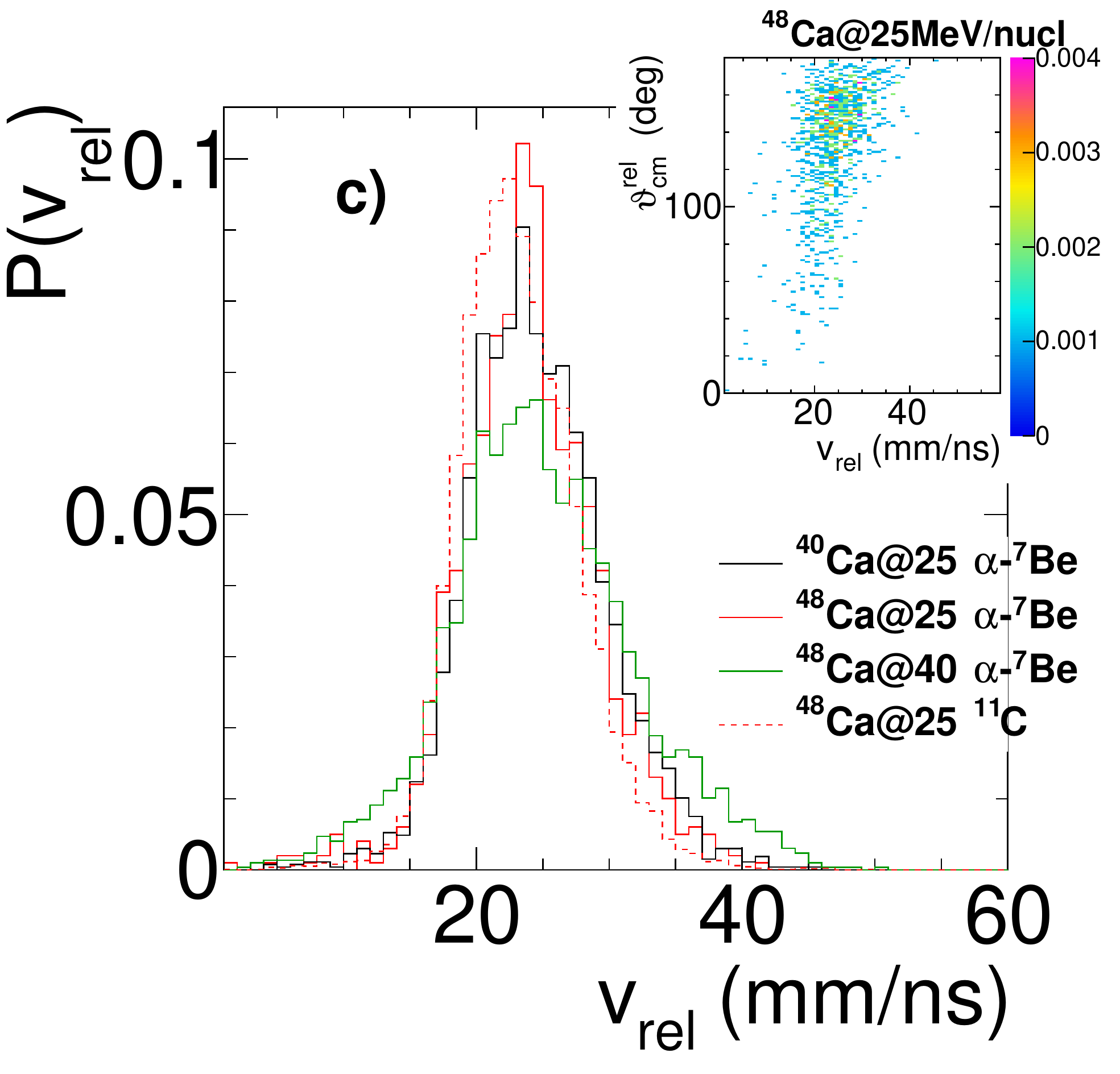}
\end{tabular}
\caption{a): Relative probability distribution of the excitation energy of $^{11}$C$^{*}$ reconstructed from 
  $\alpha$- $^7$Be correlations after background subtraction. In the
  inset the 1+R distribution and the Coulomb background for $^{48}$Ca
  at 40 MeV/nucleon are shown. b): Relative probability distribution of the c.m. parallel 
  velocity of the reconstructed $^{11}$C$^{*}$ calculated with respect
  to the velocity of the biggest fragment of the event, corrected for the correlated recoil, (for events in which a Z$\geq$10 ejectile is detected in coincidence) for
  $E^{*}_{11C}<$15 MeV. c): Continuous histograms: relative probability distribution of the relative velocity
  between the c.m. of $\alpha$ and $^7$Be and the biggest
  fragment of the 
  event, for $E^{*}_{11C}<$15 MeV. Dashed histogram: the same but for the detected $^{11}$C. In
  the inset the relative c.m angle vs. the relative velocity between the heavy fragment detected in coincidence and the c.m. of $\alpha$ and $^7$Be for the system $^{48}$Ca at 25
  MeV/nucleon is shown.}  
\label{11C}
\end{figure*}
\paragraph{{\bf $^{11}$C $^*$}}
\begin{figure}[htpb]
\includegraphics[width=0.4\textwidth]{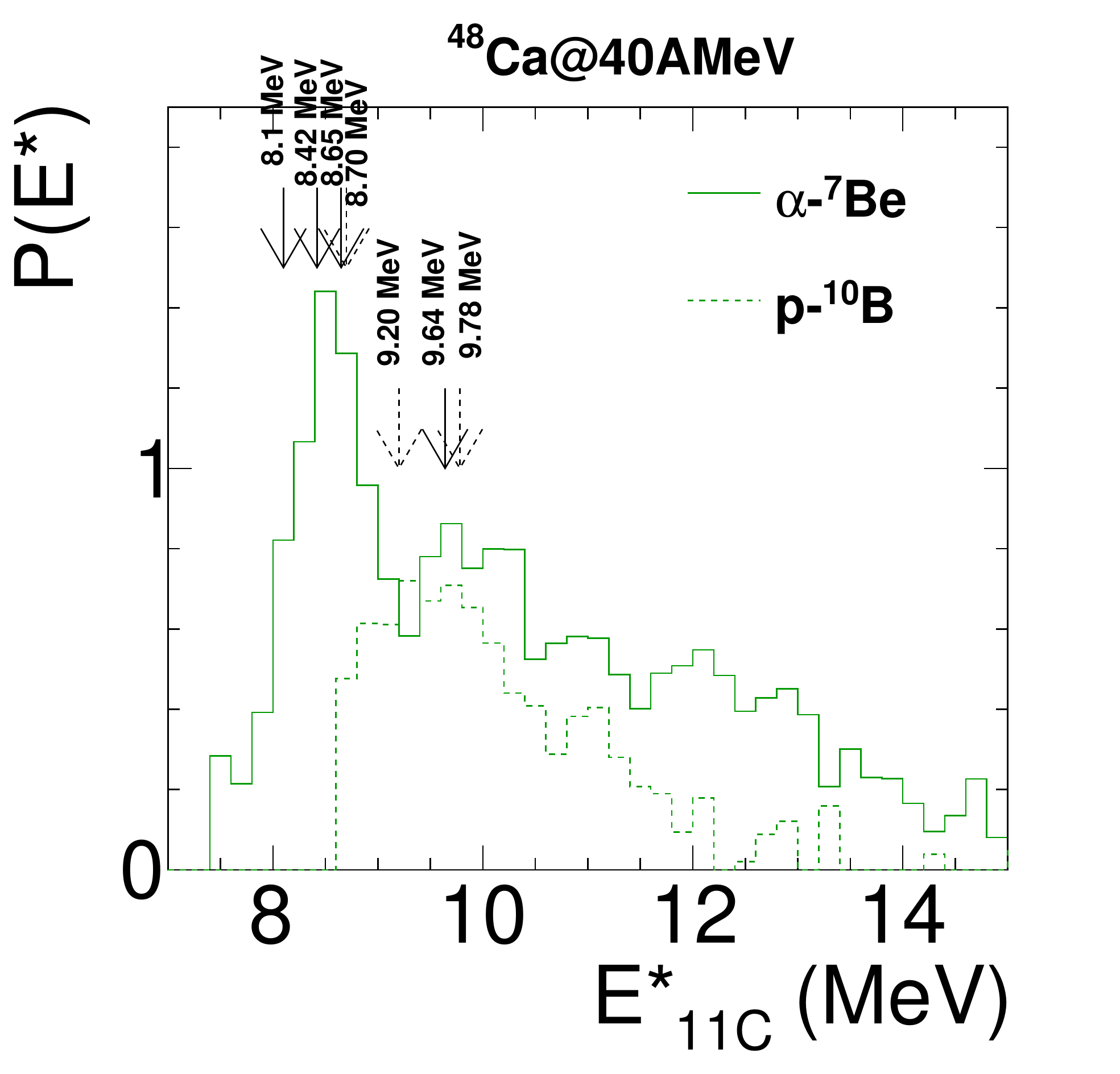}
\caption{Relative probability distributions of the excitation energy of $^{11}$C$^*$ for the reaction $^{48}$Ca
at 40 MeV/nucleon~obtained from two decay  channels; the continuous
  histogram refers to the $\alpha$+$^7$Be decay while the dashed
  histogram refers to p+$^{10}$B.  In both cases background has been
  subtracted.  Some     tabulated excited levels are marked with
  arrows (dashed arrows for    exclusive proton   decaying levels).}
\label{11C2modi}
\end{figure}
\begin{figure*}[hpt]
\begin{tabular}{ccc}
\includegraphics[width=0.33\textwidth]{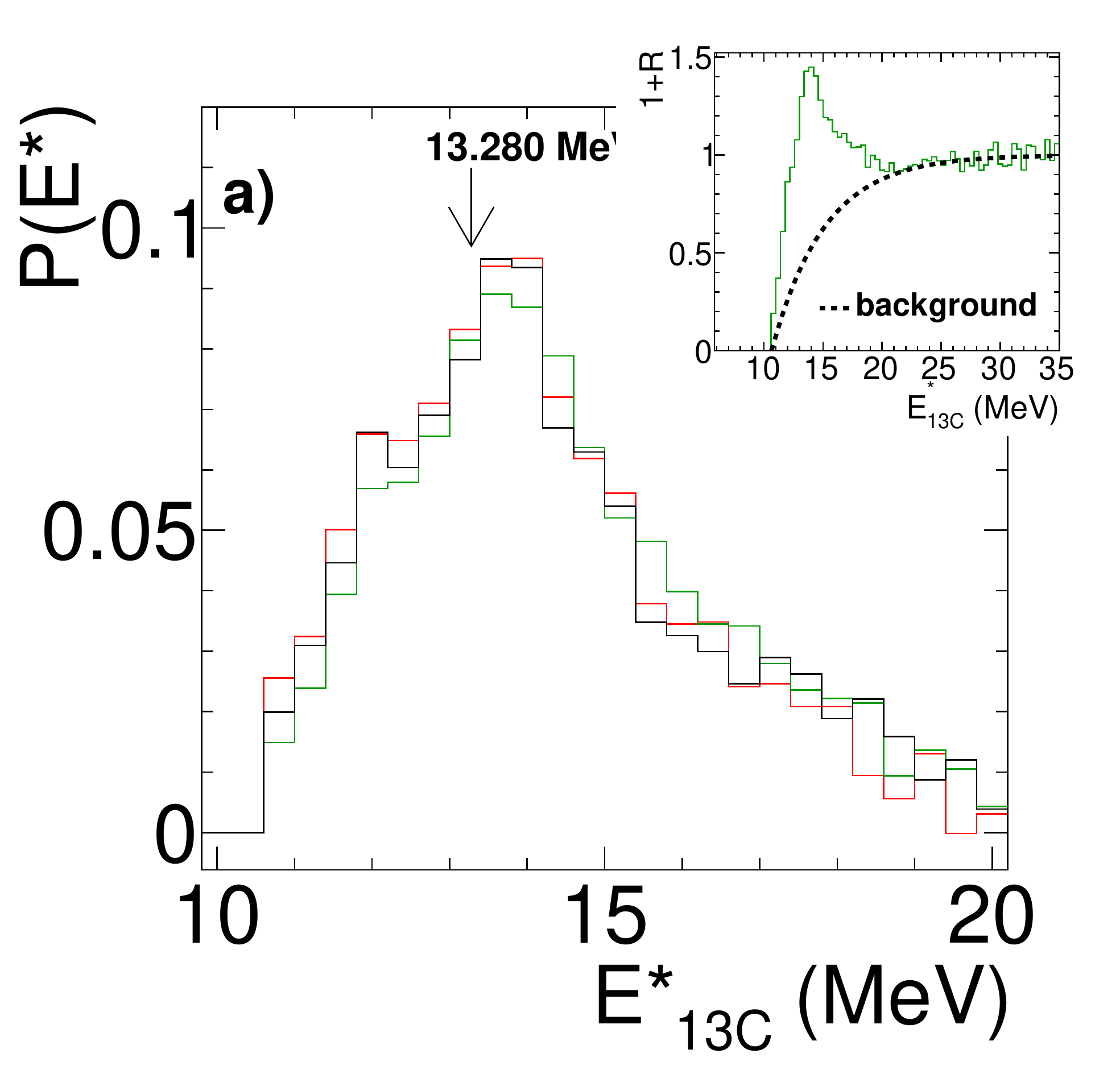}&\includegraphics[width=0.33\textwidth]{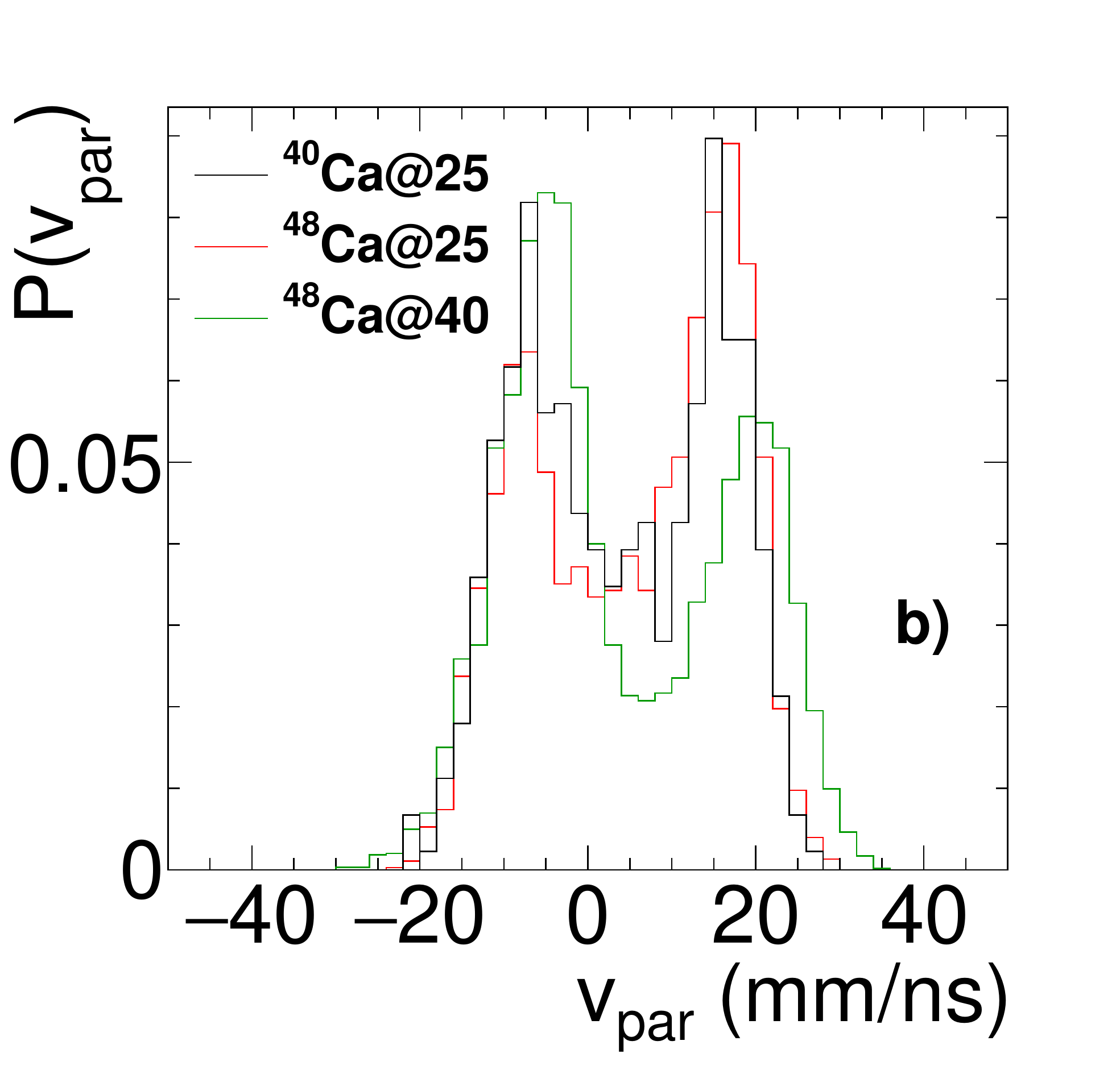}&\includegraphics[width=0.33\textwidth]{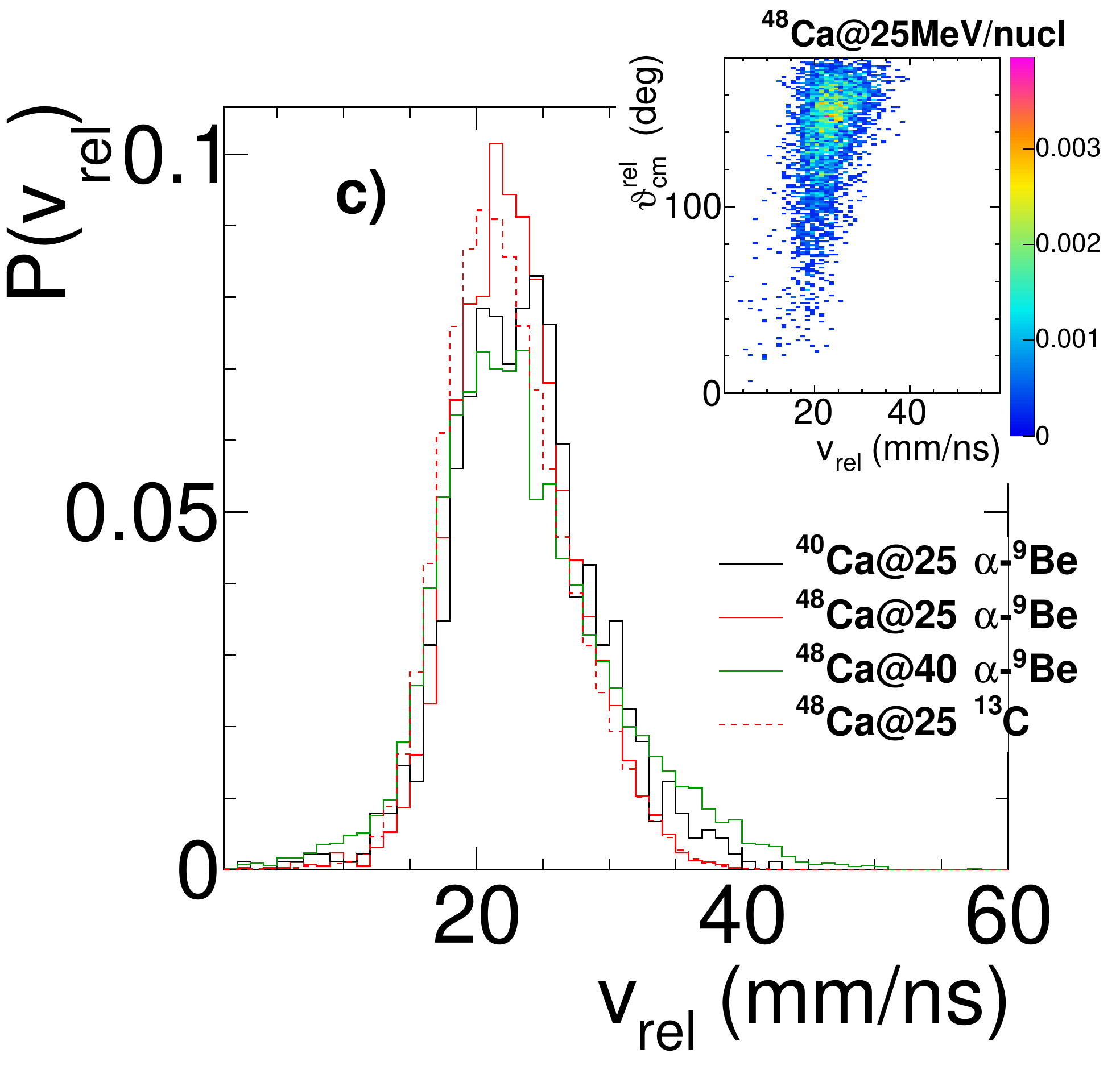}\\
\end{tabular}
\caption{a): Relative probability distribution of the excitation energy of $^{13}$C$^{*}$ reconstructed from 
  $\alpha$- $^9$Be correlations after background subtraction. In the
  inset the 1+R distribution and the Coulomb background for $^{48}$Ca
  at 40 MeV/nucleon are shown. The only tabulated level reliably
  $\alpha$ decaying is marked with an arrow. b): Relative probability distribution of the c.m. parallel 
  velocity of the reconstructed $^{13}$C$^{*}$ calculated with respect
  to the velocity of the biggest fragment of the event with Z$\geq$10, corrected for the correlated recoil, for
  $E^{*}_{13C}<$15 MeV. c): Continuous histograms: relative probability distribution of the relative velocity
  between the c.m. of $\alpha$ and $^7$Be and the biggest
  fragment of the 
  event, for $E^{*}_{13C}<$15 MeV. Dashed histogram: the same but for the detected $^{13}$C. In
  the inset the relative c.m angle vs. the relative velocity between the heavy fragment detected in coincidence and the c.m. of $\alpha$ and $^9$Be for the system $^{48}$Ca at 25
  MeV/nucleon is shown.}  
\label{13C}
\end{figure*}
The correlation technique applied to $^{12}$C can be extended to other
Carbon isotopes. In Fig. \ref{11C} panel (a) the excitation
energy spectrum of $^{11}$C$^{*}$ reconstructed from
the $\alpha$+$^{7}$Be channel, after background 
subtraction, is shown for the 3 systems; for completeness, in the
inset the 1+R 
distribution together with the estimated Coulomb background is shown
for the high energy reaction. Various structures corresponding to excited
levels can be seen in this plot; for example the first corresponds
to the level 8.654~MeV, with a 100\% probability of $\alpha$ decay,
but a contribution from the 8.42 MeV level, with a lower $\alpha$
decay probability, cannot be excluded. The peak region beyond 9.3 MeV and
below 10.5~MeV corresponds to unresolved levels, not decaying only
with $\alpha$ emission.
Within the resolution  of the present analysis and within
the statistical errors, the
populations of the different levels result to be very similar for the
3 reactions.
As a check of the reconstruction procedure, in addition to the $\alpha$-$^{7}$Be correlation, in Fig.\ref{11C2modi} we show the $^{11}$C$^*$ states obtained through the p-$^{10}$B channel (dashed histogram, 40 MeV/nucleon reaction only).
%we can 
%reconstruct $^{11}$C$^*$ states through the 
%p-$^{10}$B channel,
%as well. The result is presented in
%Fig. \ref{11C2modi} (dashed histogram) only for the 40 MeV/nucleon reaction, where the excitation energy
%spectrum is compared to that of $^{11}$C$^*$  from the
%$\alpha$-$^7$Be channel (continuous
%histogram). 
%The spectra have different lower limits due to the different Q-values (higher in absolute value for
%the
%proton channel) and because 
%the various excited levels of $^{11}$C  have
%different branching ratios for  the proton or $\alpha$ decays;
%for example the peak at 9.2~MeV could be seen only through proton
%emission, while for some levels both decay channel are allowed such as 
%the 9.6~MeV level.}
Within the limit of the present resolution we indeed
find results coherent with the known level structure.

Fig. \ref{11C} panels (b) and (c) shows the same distributions already built for the 3-$\alpha$ case (Fig. \ref{12C} (b) and (c)) for the system $\alpha$-$^{7}$Be;
the plots, normalized to the integral, are obtained requiring the coincident 
Z$\geq$10 fragment and with the gate $E^{*}_{11C}<$15 MeV.
The spectra for $^{11}$C resonances are overall very similar to those
obtained for  $^{12}$C states; also in this case the background contribution is small due to the fact that we are gating on a 3 body coincidence between $\alpha$, $^{7}$Be and a fragment with Z$\geq$10.
In conclusion,  the velocity patterns 
(panels (b) and (c)) confirm that in the examined events the $^{11}$C (detected or reconstructed) is the $LF$ of the breakup process of a heavier source.

\paragraph{\bf{$^{13}$C$^*$}}

Finally, we apply the same analysis to the n-rich $^{13}$C$^*$
resonances
showing the results in the same manner as before in Fig. \ref{13C}.
Quite similar conclusions about the breakup origin of these
resonances can be drawn by inspecting these panels.
As for the excitation spectrum (panel (a)) we note again the
remarkable similarity of the three reactions. Moreover, the details of
the excited levels of $^{13}$C$^*$ are less known \cite{nota};
% \footnote{https://www.nndc.bnl.gov/nudat3/getdataset.jsp?nucleus=13C\&unc=NDS%};
only one $\alpha$ decaying level with 100\% probability and width of 340 keV is
tabulated (around 13.28 MeV, the energy is slightly uncertain). Other
levels that are candidate for $\alpha$ decay are known, but
the probability is undefined. The reconstructed energy spectrum (after
background subtraction) for $^{13}$C$^*$ is therefore interesting and
overall compatible with the partially known structure.
Panels (b) and (c) of Fig. \ref{13C} lead to analogous
conclusions about the origin of the  $^{13}$C, either produced beyond or below the 
particle separation energy: namely, 
the findings strongly support that the emission pattern of the
reconstructed $^{13}$C$^*$ is 
compatible with a breakup process from either
an  incomplete fusion source or from very dissipative collisions.

\section{Summary and Conclusions}

In this work, experimental data for the asymmetric reactions
$^{40,48}$Ca+$^{12}$C at 25 MeV/nucleon~and $^{48}$Ca+$^{12}$C at 40 MeV/nucleon,
collected with 6 blocks of the FAZIA setup, have been discussed. The data set mainly
corresponds to both
incomplete fusion and strongly damped binary collisions which produce various primary nuclei, possibly deformed and with broad excitation
energy distributions, that may undergo a breakup process.
The analysis focused on  some features of the breakup
events  which have been investigated with unprecedented details thanks
to the excellent isotopic separation achieved with the FAZIA telescopes (up
to Z$\sim$25). After a general characterization of BU events from the point of view of the charge asymmetry, of the size of the fragments and of their isotopic composition, the relative velocity between the breakup fragments has been accurately
studied and discussed in comparison with the predictions of the low-energy 
Coulomb-driven systematics for fission, taking into account the measured charge and mass of both $HF$ and $LF$. A growth of the $v_{rel}$ to $v_{Viola}$ ratio with the beam energy was observed, perhaps related to stronger dynamical effects. 

If the light fragment coming from the BU is excited beyond particle
separation energy, it is possible to reconstruct the $LF$ by means of
the particle correlation, as it was shown for some Carbon
isotopes on exclusive many body events. To our knowledge, a similar analysis on the BU channel is not present in the literature. The obtained emission pattern and in particular the relative
velocity between the reconstructed $LF$ and the detected $HF$ turned
out to be very similar to the case in which the Carbon isotope was
detected as such.  

It is our intention to extend the analysis of the BU channel to other
available data sets. For example, the case of the light asymmetric
systems $^{32}$S,$^{20}$Ne+$^{12}$C at 25 MeV/nucleon~and 50 MeV/nucleon
\cite{Frosin2022} may allow for the investigation of  sources
produced through   very damped  collisions (as in this case)
but with smaller sizes, with possible effects on the relative velocity.
The QP
breakup will be instead investigated on the almost symmetric systems
$^{58,64}$Ni+$^{58,64}$Ni at 32 MeV/nucleon and 52 MeV/nucleon
\cite{Ciampi2022}, taking advantage of the high degree of completeness
of these data, which where collected with the INDRA FAZIA setup.

\section*{ACKNOWLEDGMENTS} 
We would like to thank the accelerator staff of LNS laboratories for having provided good-quality beams and support during the experiment.
We sincerely thank A.Ono for the use of the AMD code.
This work was partially funded by the Spanish Ministerio de Econom\'ia y Empresa (PGC2018-096994-B-C22)
\bibliography{biblio}

\end{document}